\documentclass[final,authoryear,5p,times,twocolumn]{elsarticle}

\usepackage{graphicx}
\usepackage{amssymb}
\usepackage{amsthm}
\usepackage{amsmath} 
\usepackage{amsfonts} 
\usepackage{array} 
\usepackage[boxed]{algorithm2e}

\biboptions{longnamesfirst}

\journal{Expert Systems with Applications}

\begin{document}
\begin{frontmatter}

\title{A data mining approach using transaction patterns for card fraud detection\tnoteref{label1}}
\tnotetext[label1]{This research was supported by the MKE (Ministry of Knowledge Economy), Korea, under the ``Employment Contract based Master's Degree Program for Information Security" supervised by the KISA (Korea Internet Security Agency)}

\author{Chae Chang Lee}
\ead{chiching@korea.ac.kr}

\author{Ji Won Yoon\corref{cor1}}
\ead{jiwon\_yoon@korea.ac.kr}
\ead[url]{https://sites.google.com/site/securesiplab/}

\cortext[cor1]{Corresponding author. Tel:+82 2 3290 4886}
\address{Graduate School of Information Security, Korea University, Seoul, Republic of Korea}


\begin{abstract}

Credit and debit cards, rather than actual money, have become the universal payment means. With these cards, it has become possible to buy expensive items easily without an additional complex authentication procedure being conducted. However, card transaction features are targeted by criminals seeking to use a lost or stolen card and looking for a chance to replicate it. Accidents, whether caused by the negligence of users or not, that lead to a transaction being performed by a criminal rather than the authorized card user should be prevented. Therefore, card companies are providing their clients with a variety of policies and standards to cover this eventuality. Card companies must therefore be able to distinguish between the rightful user and illegal users according to these standards in order to minimize damage resulting from unauthorized transactions.

However, there is a limit to applying the same fixed standards to all card users, since the transaction patterns of people differ and even individuals' transaction patterns may change frequently due to changes income and consumption preference. Therefore, when only a specific threshold is applied, it is difficult to distinguish a fraudulent card transaction from a legitimate one.

In this paper, we present methods for learning the individual patterns of a card user's transaction amount and the region in which he or she uses the card, for a given period, and for determining whether the specified transaction is allowable in accordance with these learned user transaction patterns. Then, we classify legitimate transactions and fraudulent transactions by setting thresholds based on the learned individual patterns.

\end{abstract}


\end{frontmatter}




\section{Introduction}
\label{section:Sec1}
A lot of crime involving card fraud is being perpetrated due to transaction card cloning or theft. Thus, in order to prevent fraudulent transactions due the replication of a card at source, financial authorities in many countries are moving from the Magnetic Stripe (MS) card to the Integrated Circuit (IC) card. However in contrast to banks which can easily introduce ATMs dedicated to IC cards, it is not easy for retail stores to replace the previous card reader at Point of Sale (POS) terminals with an IC reader device. Therefore, at present, the card data, which include the consumer's credit information and transaction data are included in both the MS and IC of a card, and as a result, it seems still difficult to prevent illegal card use resulting from theft or replication.

Card companies currently have introduced a variety of measures to prevent damage being caused to the consumer by an unauthorized card transaction, such as sending an SMS about the transaction and a copy of the transaction by E-mail, suspending the use of the credit card, and so on. However, these methods are dependent on consumer's attention. When they do not pay attention to the received billing messages and are therefore not aware of a fraudulent transaction perpetrated using their card, it is difficult to detect fraud cases early and take appropriate action.

The card companies may take voluntary action to prevent unauthorized payments by determining in what location and the amount of the transaction for which a card is used. For example, if a payment involving a large amount of money is made in Southeast Asia, and the previous transaction had occurred on hour previously in Seoul, Republic of Korea, the company may conduct a verification process through a phone call to the customer. Because the same card cannot be used almost simultaneously in two distant locations, a rational explanation is that a duplicate of the card was used.

However, there is a limit to the number of fraudulent transactions that can be detected by simply using a fixed threshold value for the geographic distance between transactions, and the location or amount of payment, because the consumption levels of people differ, and a cloned or lost card can be used in a location close to the legitimate owner's normal use area.

In this paper, we therefore propose methods for detecting an illegal card transaction using large scale data analysis. The customers' unique billing pattern is used in the process. In the next section, we provide related studies and discuss the differences between these and our study. In section 3, we provide the background of statistical methods and data mining approaches. In section 4, we discuss the model for our experiment on the detection of fraudulent transactions, which is based on transaction patterns. In section 5, we present our experimental results, and we discuss the reasons for the results in section 6. Finally, we present our conclusions in section 7.


\section{Material and methods}
\label{section:Sec2}
Data mining is finding patterns in data that are statistically reliable, previously unknown, and can be analyzed to provide useful insights \citep{ref1}. Fraud detection is a main research area in the field of data mining. The goal of using the data mining approach in the detection of card fraud is to determine whether the card used in a transaction has been used by the legitimate card user. Here, a card is a means of payment used for transactions, such as a credit, debit, or purchase card, and the user is the authorized owner of the card. The data used for data mining constitutes a user's transaction records.

Statistical fraud detection methods have been classified into two broad categories: ``supervised" and ``unsupervised" \citep{ref2, ref3}. In supervised methods, estimated statistical models are used to discriminate between fraudulent and legitimate purchase behaviors by classifying new observations into the appropriate class: fraudulent or legitimate transaction \citep{ref2, ref3}. This method requires samples from both classes, fraudulent and legitimate observations, as the models are trained based on examples of observations in both classes \citep{ref3}. The models created by the method are assessed by measuring their accuracy in correctly classifying new observations as fraudulent or non-fraudulent \citep{ref3}. Since 2001, most fraud detection studies using supervised algorithms have focused on the misclassification rate (i.e., the false positive and false negative error rate) \citep{ref4}.

Related works in which supervised methods are used for classifying credit card transactions into legitimate and fraudulent transactions have been published: \citet{ref3} employed a transaction aggregation strategy to create variables for the estimation of a logistic regression model, and \citet{ref6} used a binary support vector system based on the support vectors in support vector machines and a genetic algorithm to solve credit card fraud problems that had not been well identified \citep{ref5}.

Unsupervised methods attempt to detect unusual observations, such as customers, transactions, or accounts whose behavior may be different from the norm \citep{ref3}, which can be identified by clustering based on normal legitimate behavior patterns. As unsupervised methods do not require samples of fraudulent and legitimate transactions, there may be cases where there is no prior knowledge of classes of observations \citep{ref3}.

In earlier studies, unsupervised methods were also used for clustering to detect credit card fraud. \citet{ref7} built a Hidden Markov Model for the sequence of operations in credit card transaction processing. \citet{ref8} focused on real-time fraud detection and presented a new model. \citet{ref9} created a model of typical cardholder behavior and analyzed deviations in transactions in order to identify suspicious ones. The studies of \citet{ref8} and \citet{ref9} were based on a self-organizing map algorithm \citep{ref5}.

A data mining approach for mobility patterns is also used to predict the location of drivers or mobile phone users. \citet{ref12} presented a Hidden Markov Model-based approach to provide real-time predictions of a driver's destination and route. \citet{ref13} focused on the regions of interest and the typical travel time of moving objects from region to region. They introduced a novel form of spatio-temporal pattern, which formalizes the idea of aggregate movement behavior which they discuss. \citet{ref15} proposed a context model, based on classification using a certain moving profile and history of movements. They evaluated their schemes with Bayesian algorithms, decision-tree, and rule induction.

Using supervised methods to detect fraudulent card transaction involves many constraints. Because academicians have difficulty acquiring credit card transaction datasets, it is not easy to exchange ideas and possible innovations related to credit card fraud detection because of the dearth of published literature in this subject \citep{ref10, ref11, ref3}.

In this paper, we therefore detect fraudulent transaction based on the pattern of the amount of a customer's payment using the purchase card data that are publicly available. The pattern is analyzed through Gaussian Processes (GPs) and the Autoregressive (AR) model, which have not previously been well documented in the area of fraud detection. Further, to characterize the pattern of a customer's payment region, we collect the location of the customer and track the mobility confidence. Then, we find abnormal values by extracting the association rules of the payment region of a card user by applying an association rule algorithm \citep{ref16}, which has previously been used to predict the movement of mobile phone users.


\section{Theory and calculation}
\label{section:Sec3}


\subsection{Autoregressive model}
\label{section:Sec3.1}
Given a dataset of $N$ observations $\left \{ y_i \right \}_{i=1}^{N}$, where $y_i\in \mathbb{R}$, $AR(p)$ is the autoregressive model that expresses the $i$-th data with $p$ prior data $\left \{ y_j \right \}_{j=i-1}^{i-p}$. The equation of $AR(p)$ is defined as
\[
y_i=a_0+\sum_{j=1}^p a_jy_{i-j}+\epsilon_i
\]
where $\left \{ a_j \right \}_{j=1}^{p}$ are parameters of the model, $a_0$ is a constant, and $\epsilon_i$ is white noise.

We can express the terms from $y_{1}$ to $y_{N}$ by
\begin{equation} \label{eq:matrix}
\begin{pmatrix}
y_{1}\\ y_{2}\\ y_3 \\ \vdots\\ y_N
\end{pmatrix}
=
\begin{pmatrix} 
1 & 0 & 0 & \ldots & 0 \\ 
1 & y_1 & 0 & \ldots & 0 \\ 
1 & y_2 & y_1 & \ldots & 0\\ 
\vdots & \vdots & \vdots & \vdots & \vdots \\ 
1 & y_{N-1} & y_{N-2} & \ldots & y_{N-p}
\end{pmatrix}
\begin{pmatrix}
a_0\\a_1\\a_2\\ \vdots\\ a_p
\end{pmatrix}
+ \epsilon.
\end{equation}

Here, $y_i$, where $1 \leq i \leq N$, are training data and used to estimate a target $y_{N+1}$. In order to impose a restriction such that not all previous data can be used to infer the target data, it is also possible to utilize only the data close to the target data by using a fixed small window size. Let the left hand term of equation \eqref{eq:matrix} be $Y$, the $N \times (p+1)$ matrix of the right term be $X$, and the matrix that contains unknown values $a_0,a_1,\cdots,a_n$, be $A$. Then, the equation can be rewritten as a simple linear form given by
\begin{equation} \label{eq:linear}
Y=XA+\epsilon.
\end{equation}

This represents that equation \eqref{eq:linear} is a simple linear model with input data $X$ and parameters $A$. The observed data $Y$ have a noise term $\epsilon$, which is assumed to be independent and identically distributed (i.i.d.) Gaussian distribution
\[
\epsilon \sim \mathcal{N}(0, \tau^{-1}I),
\]
where $\mathcal{N}(a, b)$ denotes a Gaussian distribution with mean $a$ and variance $b$, and $\tau$ is a precision value that is an inverse of the variance.

This noise assumption, together with the model, directly gives rise to the likelihood (i.e., the probability density of the observations given the parameters), which is factored over the cases in the training set (because of the independence assumption) to give \citep[chap. 2]{Gaussian06}
\begin{eqnarray} \label{eq:P_distribution}
P(Y|X,A) &=&
\prod_{i=1}^Np(y_i|x_i,A)\nonumber \\
&=&\prod _{i=1}^N\frac{1}{(2\pi\tau^{-1})^{1/2}}\exp\left \{-\frac{(y_i-x_iA)^2}{2\tau^{-1}}\right \}\nonumber\\
&=&\frac{1}{(2\pi\tau^{-1})^{N/2}}\exp\left \{-\frac{1}{2\tau^{-1}} \left | Y-XA \right |^2 \right\}\nonumber\\
&=&\mathcal{N}(XA,\tau^{-1}I).
\end{eqnarray}

Then, the residual vector is $Y-XA$. A generalized least squares method is used to estimate $A$ by minimizing the squared Mahalanobis length of this residual vector:
\[
\hat{A} = \arg_{A}\min\left \{ (Y-XA)^T\tau^{-1}I(Y-XA) \right \}.
\]
Hence,
\begin{equation} \label{eq:hat_A}
\hat{A} = (X^TX)^{-1}X^TY.
\end{equation}

The variance of an error term $\epsilon$, $\tau^{-1}$, is obtained by the maximum likelihood
\[
\hat{\tau}^{-1} = \arg_{\tau^{-1}}\max P(Y|\hat{A},\tau^{-1}).
\]
Substituting $P(Y|\hat{A},\tau^{-1})$ for the form of the Gaussian distribution, we obtain the log likelihood function in the form \citep[chap. 1]{bishop07}
\begin{eqnarray} \label{eq:logarithm}
\mathcal{L}&=&\ln P(Y|\hat{A},\tau^{-1}) \nonumber \\
&=&-\frac{\tau}{2}(Y-X\hat{A})^T(Y-X\hat{A})+\frac{N}{2}\ln\frac{\tau}{2\pi}.
\end{eqnarray}
Now, we can obtain the estimated precision by inducing the differential equation of equation \eqref{eq:logarithm} equal to 0 as
\begin{equation} \label{eq:tau}
\tau^{-1}=\frac{1}{N}(Y-X\hat{A})^T(Y-X\hat{A}).
\end{equation}

Now, based on the probability distribution of equation \eqref{eq:P_distribution}, the target data $y_{N+1}$ can be predicted by expectation $\hat{Y}=X\hat{A}$ with a $95\%$ confidence level with the range:
\[
\hat{Y}-2\tau^{-1}  \leq y_{N+1} \leq \hat{Y}+2\tau^{-1}.
\]


\subsection{Gaussian Processes}
\label{section:Sec3.2}
Gaussian Processes (GPs) are the extension of multivariate Gaussians to infinite-sized collections of real-valued variables and distributions over random functions \citep{GP08}. That is, they predict the subsequent data by revealing the distribution of the nonlinear function $\mathbf{f}=\left \{ f_i \right \}_{i=1}^N$, which represents the relationship between the output data $\mathbf{y}=\left \{ y_i \right \}_{i=1}^N$ and the input data $\mathbf{x}=\left \{ x_i \right \}_{i=1}^N$.

Each observation $y\in \mathbb{R}$ from its corresponding input data $x$ is given by
\[
y=f(x)+ \epsilon
\]
through the Gaussian noise model with variance $\sigma_n^2$. The function $f(x)$ can be expressed as $\mathbf{f}\sim \mathcal{N}(\mathbf{m},\mathbf{k})$ given by
\[
\begin{bmatrix}
\mathbf{f}(x_1)\\ 
\vdots\\ 
\mathbf{f}(x_N)
\end{bmatrix}
\sim \mathcal{N}
\left ( 
\begin{bmatrix}
\mathbf{m}(x_1)\\ 
\vdots\\ 
\mathbf{m}(x_N)
\end{bmatrix},
\begin{bmatrix}
\mathbf{k}(x_1,x_1) & \cdots & \mathbf{k}(x_1,x_N)\\ 
\vdots & \ddots & \vdots\\ 
\mathbf{k}(x_N,x_1) & \cdots & \mathbf{k}(x_N,x_N)
\end{bmatrix} \right ),
\]
where $\mathbf{m}$ is the mean function and $\mathbf{k}$ is the covariance function \citep{GP08}. Usually, for notational simplicity we take the mean function to be zero. We now choose the covariance function by writing
\begin{equation} \label{eq:20}
k(x,{x}')=\sigma_f^2\exp\left \{ \frac{-(x-{x}')^2}{2l^2} \right \}+\sigma_n^2\delta(x,{x}'),
\end{equation} 
where $\delta(x,{x}')$is the Kronecher delta function \citep{Gaussian06}. Its parameters $\mathbf{\theta}=\left \{ l,\sigma_f,\sigma_n \right \}$ can be estimated through Bayes' theorem. According to Bayes and \citet{Gaussian06}, when we have little prior knowledge about what $\mathbf{\theta}$ should be, this corresponds to maximizing $\ln p(\mathbf{y}|\mathbf{x},\mathbf{\theta})$, given by
\begin{equation} \label{eq:21}
\ln p(\mathbf y|\mathbf x,\mathbf \theta)=-\frac{1}{2}\mathbf y^TK^{-1}\mathbf y-\frac{1}{2}\ln\left | K \right |-\frac{n}{2}\ln2\pi.
\end{equation}
For estimating the parameters, the Nelder-Mead simplex, multivariate optimization algorithm \citep{ref20} is one method that can be used.

However, in this study it is not enough to determine the parameters using equation \eqref{eq:21}. Because there is no restriction on parameter $l$, it can have any value from a negative value to $10^8$. In this case, we use maximum a posteriori (MAP) to find parameter $l$ as
\begin{eqnarray} \label{eq:map}
\ln p(l|\mathbf{y})&=&\ln\frac{p(\mathbf{y}|l)p(l)}{p(\mathbf{y})} \nonumber \\
&=& \ln p(\mathbf{y}|l) + \ln p(l).
\end{eqnarray} 
We can replace the term $\ln p(\mathbf{y}|l)$ in equation \eqref{eq:map} with equation \eqref{eq:21}. In addition, we need to constrain parameter $l$ in order to interpret testing data $y_{N+1}$ as an event related to the close training data. That is, to estimate the 101-st testing data we focus more on the 100-th training data than on the 50-th. We thus assume that the parameter $l$ follows a non-negative distribution $l \sim \Gamma (2,2)$, where it is the gamma distribution with a shape parameter 2 and a scale parameter 2. It is given by
\begin{equation} \label{eq:para_theta}
\ln p(\theta|\mathbf{y})=-\frac{1}{2}\mathbf y^TK^{-1}\mathbf y-\frac{1}{2}\ln\left | K \right |-\frac{n}{2}\ln2\pi + \ln p(l).
\end{equation}
Hence, we can obtain restricted parameter $l$, as well as $\sigma_f$ and $\sigma_n$, by calculating the arguments that maximize equation \eqref{eq:para_theta}. For simplicity, we assumed the prior distribution of $\sigma_f$ and $\sigma_n$ follow uniform distribution, which is an improper distribution.

Now, given $N$ observations of $\mathbf{y}$, in order to predict not the actual $f_*$ but $y_*$, we prepare three matrices.
\begin{eqnarray}
K&=&
\begin{bmatrix}
k(x_1,x_1) & k(x_1,x_2) & \cdots & k(x_1,x_N)\\ 
k(x_2,x_1) & k(x_2,x_2) & \cdots & k(x_2,x_N)\\ 
\vdots & \vdots & \ddots & \vdots\\ 
k(x_N,x_1) & k(x_N,x_2) & \cdots & k(x_N,x_N)
\end{bmatrix} \label{eq:22}
\\
K_*&=&
\begin{bmatrix}
k(x_*,x_1) & k(x_*,x_2) & \cdots & k(x_*,x_N) 
\end{bmatrix} \label{eq:23}
\\
K_{**}&=&k(x_*,x_*). \label{eq:24}
\end{eqnarray} 

Then, $y_*$ can be predicted based on $N$ observations of $\mathbf{y}$ as a sample from a multivariate Gaussian distribution:
\[
\begin{bmatrix}
\mathbf{y}\\ 
y_*
\end{bmatrix}
\sim
\mathcal{N}
\left ( \mathbf{0},
\begin{bmatrix}
K & K_*^T\\ 
K_* & K_{**}
\end{bmatrix} \right )
\]
Given the data, a certain prediction for $y_*$ can be obtained by conditional Gaussian distributions as
\[
y_*|\mathbf{y}\sim \mathcal{N}(K_*K^{-1}\mathbf{y}, K_{**}-K_*K^{-1}K_*^T).
\]
That is, the best estimate for $y_*$ follows the distribution that has the expectation of $y_*$, $\overline{y}_*=K_*K^{-1}\mathbf{y}$, and the uncertainty, $var(y_*)=K_{**}-K_*K^{-1}K_*^T$. Therefore, $y_*$is decided with the range:
\[
\overline{y}_*-2\sqrt{var(y_*)} \leq y_* \leq \overline{y}_*+2\sqrt{var(y_*)},
\]
giving a 95\% confidence level.


\subsection{Extreme-value theory}
\label{section:Sec3.3}
Extreme-value theory is a branch of statistics that concerns the distributions of data of unusually low or high value \citep{ref23}. It forms representations for the tails of distributions, in this article especially the right-hand tail.

Assume a set $Z_m=\left \{ z_1,z_2,\cdots,z_m \right \}$ with $m$ i.i.d. random variables drawn from the one-sided standard Gaussian (i.e., $\mathcal{D}=\left | \mathcal{N}(0,1) \right |$ ) and define $z=\max(Z_m)$, which means the largest element observed in $m$ samples of $Z_m$. Then, the extreme value probability (EVP) of $z$ is its probability of  being the largest value in the set and is obtained in the form of the Gumbel distribution as \citep{ref23, ref24}
\begin{equation}
P_{EV}(z|m)=\exp\left \{ -\exp\left ( -\frac{z-\mu_m}{\sigma_m} \right ) \right \},
\end{equation}
where the location parameter is
\begin{equation}
\mu_m=(2\ln m)^{1/2}-\frac{\ln(\ln m)+\ln 2\pi}{2(2\ln m)^{1/2}},
\end{equation}
and the scale parameter is
\begin{equation}
\sigma_m=(2\ln m)^{-1/2}.
\end{equation}

Since the parameters $\mu_m$ and $\sigma_m$ depend on the number of data, $m$, $m$ should be specified first in order to evaluate an EVP. However, because we are not directly interested in $m$, we obtain the EVP at time $t$ by marginalizing out the run length $l_t$ \citep{ref24}:
\begin{equation} \label{eq:EVP}
P_{EV}(t)=\sum_{m=1}^{t}P_{EV}(t|l_{t}=m)P(l_{t}=m).
\end{equation}
The run length $l_t$ means the time since the last outlier \citep{ref25}, and its probability \citep{ref24} is 
\[
P(l_t=m)=
\begin{cases}
P_{EV}(t-1) & \text{ if } m=1 \\ 
\left ( 1-P_{EV}(t-1) \right )P(l_{t-1}=m-1) & \text{ if } m \geq 2,
\end{cases}
\]
where $P(l_t=1)=1$. Then, the EVP can be used as a novelty measure and an outlier can be detected in the case of $P(y)>\theta_{EV}$, where $\theta_{EV}$ is a threshold with $0 \leq \theta_{EV} \leq 1$ \citep{ref24}.


\subsection{Association rule}
\label{section:Sec3.4}
Association rule analysis is a data mining technique used to find which events are likely to co-occur. In this study, it is used to extract the region where transactions occur frequently and the pattern of the movement path, by indicating the transaction locations as a sequence. For notational convenience, $\left \langle a^{(i)},b^{(j)} \right \rangle$ defines the path sequence of a customer who made payments with a card $i$ times in region $a$ to $j$ times in region $b$. For instance, a path sequence $\left \langle 1, 2^{(2)}, 3 \right \rangle=\left \langle 1,2,2,3 \right \rangle$ represents the transition of a customer's location, such as, $1\rightarrow 2\rightarrow 2\rightarrow 3$. Table \ref{tab:T1} is an example that shows the route where the transactions of a certain customer occurred during a month.
\begin{table}[h]
\begin{center}
    \begin{tabular}{c|c}
    Week(s) ago & Actual transaction path sequence                            \\ \hline
    4           & $\left \langle 7,1^{(1)},2 \right \rangle$                  \\
    3           & $\left \langle 6^{(2)},9,4^{(3)},10,1^{(2)} \right \rangle$ \\
    2           & $\left \langle 1^{(3)},6^{(2)},1,12,3 \right \rangle$       \\
    1           & $\left \langle 8,11 \right \rangle$                         \\
    \end{tabular}
    \caption {An example of four sequences of an actual transaction path}
    \label{tab:T1}
\end{center}
\end{table}

By padding the end of each row of Table \ref{tab:T1} with 0's so that the rows have the same number of columns, we obtain the $4 \times 9$ matrix of the collected transaction path, $\mathbf{U}$, as
\begin{equation} \label{eq:33}
\mathbf{U}=
\begin{pmatrix}
7 & 1 & 1 & 2 & 0 & 0 & 0 & 0 & 0\\ 
6 & 6 & 9 & 4 & 4 & 4 & 10 & 1 & 1\\ 
1 & 1 & 1 & 6 & 6 & 1 & 12 & 3 & 0\\ 
8 & 11 & 0 & 0 & 0 & 0 & 0 & 0 & 0
\end{pmatrix}.
\end{equation}

We can now infer which regions (i.e., numbers) are likely to follow each other and which regions (i.e., numbers) are associated with others based on each row of $\mathbf{U}$. Let $R(\mathbf{U})$ be a collection of row spaces $\left \{ \mathbf{U}_i \right \}_{i=1}^4$ of matrix $\mathbf{U}$. By applying the association rule to the row spaces, $R(\mathbf{U})$, movement patterns are mined using the support value by discovering the correlation between the regions where a card user made payments.

From Table \ref{tab:T1}, the supports of the length-$1$ pattern, $P_1$, are determined by the number of row spaces that contain the pattern's element. The support value of $\left \langle a \right \rangle$, an element of $P_1$, is calculated by
\begin{equation} \label{eq:S1}
S_{\left \langle a \right \rangle} =
\left | \left \{ R(U)|\left \langle a \right \rangle \subset R(\mathbf{U}) \right \} \right |.
\end{equation}
 
The supports of the length-$n$ pattern, $P_n$, are obtained from the sum of the number of row spaces that contain the length-$n$ sequences and an incident support, $S_{INC}$:
\begin{equation}
S_{\left \langle a_{1}, \cdots ,a_{n} \right \rangle}
=\left | \left \{ R(\mathbf{U})|\left \langle a_{1}, \cdots ,a_{n} \right \rangle \subset R(\mathbf{U}) \right \} \right |+S_{INC}.
\end{equation}
The incident support is a value that reflects the region where the customer's transactions diverged from the pattern of arriving at $a_n$ along the original path (i.e., $a_2,\cdots,a_{n-1}$) from $a_1$, and is given by
\begin{equation} \label{eq:36}
S_{INC}=\begin{cases}
\frac{1}{1+t} & \text{, in case of}\left \langle a_1,b_1,\cdots,b_t,a_n   \right \rangle  \\
0 & \text{, otherwise}.
\end{cases}
\end{equation}
That is, the incident support plays the role of improving the supports by correcting the result value, even when a customer goes indirectly to the destination moving along a slightly different path from that which was mined. For instance, the support value for a pattern $\left \langle 6,1 \right \rangle$ based on $\mathbf{U}$ is $S_{\left \langle 6,1 \right \rangle} = 1+\frac{1}{1+5} =1.1667$, since for $\mathbf{U}_2$, the second row space of $\mathbf{U}$, $\left \langle 6,\dot9,\dot4,\dot4,\dot4,\dot{10},1 \right \rangle \subset \mathbf{U}_2$ and for $\mathbf{U}_3$, the third row space of $\mathbf{U}$, $\left \langle 6,1 \right \rangle \subset \mathbf{U}_3$.

\begin{table}[h]
\begin{center}
    \begin{tabular}{|cc|cccc|}
    \hline
    $P_1$                              & Support & $P_2$                                & Support & $P_2$                                   & Support \\ \hline
    $\left \langle 1 \right \rangle $  & 3       & $\left \langle 1, 1 \right \rangle  $  & 3       & $\left \langle 6, 6 \right \rangle $  & 2       \\
    $\left \langle 2 \right \rangle $  & 1       & $\left \langle 1, 2 \right \rangle  $  & 1       & $\left \langle 6, 9 \right \rangle $  & 1       \\
    $\left \langle 3 \right \rangle $  & 1       & $\left \langle 1, 3 \right \rangle  $  & 0.5     & $\left \langle 6, 10 \right \rangle $ & 0.2     \\
    $\left \langle 4 \right \rangle $  & 1       & $\left \langle 1, 6 \right \rangle  $  & 1       & $\left \langle 6, 12 \right \rangle $ & 0.5     \\
    $\left \langle 6 \right \rangle $  & 2       & $\left \langle 1, 12 \right \rangle $  & 1       & $\left \langle 7, 1 \right \rangle $  & 1       \\
    $\left \langle 7 \right \rangle $  & 1       & $\left \langle 4, 1 \right \rangle  $  & 0.5     & $\left \langle 7, 2 \right \rangle $  & 0.33    \\
    $\left \langle 8 \right \rangle $  & 1       & $\left \langle 4, 4 \right \rangle  $  & 1       & $\left \langle 8, 11 \right \rangle $ & 1       \\
    $\left \langle 9 \right \rangle $  & 1       & $\left \langle 4, 10 \right \rangle $  & 1       & $\left \langle 9, 1 \right \rangle $  & 0.2     \\
    $\left \langle 10 \right \rangle $ & 1       & $\left \langle 6, 1 \right \rangle  $  & 1.17    & $\left \langle 9, 4 \right \rangle $  & 1       \\
    $\left \langle 11 \right \rangle $ & 1       & $\left \langle 6, 3 \right \rangle  $  & 0.33    & $\left \langle 9, 10 \right \rangle $ & 0.25    \\
    $\left \langle 12 \right \rangle $ & 1       & $\left \langle 6, 4 \right \rangle  $  & 0.5     & $\left \langle 10, 1 \right \rangle $ & 1       \\
    ~                                    & ~       & ~                                      & ~     & $\left \langle 12, 3 \right \rangle $ & 1       \\ \hline
    \end{tabular}
    \caption {The patterns of length-1 and -2, and supports}
    \label{tab:support}
\end{center}
\end{table}

In Table \ref{tab:support}, the support values of length-$1$ patterns $P_1$ and length-$2$ patterns $P_2$ are given. The supports of $P_3, P_3, \cdots, P_N$ can be obtained likewise.

Now, if a card payment in a particular region has occurred, it is possible to calculate the confidence by analyzing the association between the movement path of the current consumer and the patterns of the locations of payments that were made in the past. For an association rule of transaction region $\mathrm{R}:\left \langle a_1,a_2,\cdots,a_{i-1} \right \rangle\rightarrow \left \langle a_i,a_{i+1},\cdots,a_n \right \rangle$, the confidence is given by
\begin{equation}
Conf(\mathrm{R})=
\frac{S_{\left \langle a_1,a_2,\cdots,a_n \right \rangle}}
{S_{\left \langle a_1,a_2,\cdots,a_{i-1} \right \rangle}}
\times 100.
\end{equation}

For example, suppose that we have recent transaction data where the region is $\left \langle 6,4,1 \right \rangle$ and the card owner is currently in location number 1. Then, there are three possible association rules: 
\begin{align*}
\mathrm{R}_1&:\left \langle 6 \right \rangle\rightarrow \left \langle 4,1 \right \rangle  \\
\mathrm{R}_2&:\left \langle 6,4 \right \rangle\rightarrow \left \langle 1 \right \rangle  \\
\mathrm{R}_3&:\left \langle 4 \right \rangle\rightarrow \left \langle 1 \right \rangle. 
\end{align*}
The confidence values of each rule are $Conf(\mathrm{R}_1)=10$, $Conf(\mathrm{R}_2)=40$, and $Conf(\mathrm{R}_3)=50$, and the maximum value of these results, $Conf(\mathrm{R}_3)=50$, is kept as the confidence score of the transaction path $\left \langle 6,4,1 \right \rangle$.


\subsection{Adjacency matrix}
\label{section:Sec3.5}

\begin{figure}[b!]
\includegraphics[scale=0.6]{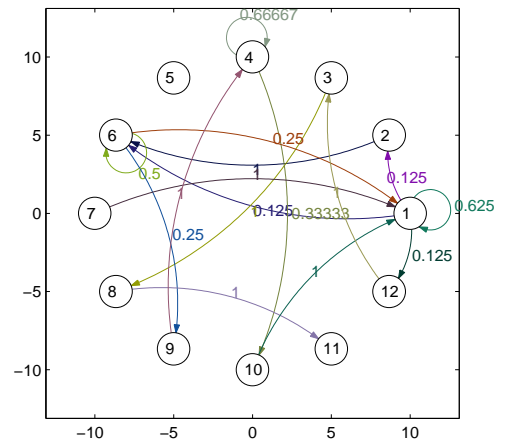}
\caption{Directed graph with conditional probability. Each node is represented by a circle, which contains the region number, and edges are represented by arrows. An arrow from one region to another means that there was a history of movement between these regions with specific conditional probability.}
\label{fig:F1}
\end{figure}

An adjacency matrix is a means of representing which vertices (or nodes) of a graph are adjacent to which other vertices \citep{wiki:AD}.
We set the transaction region, $a_i$ $(0 \leq i \leq N)$, as the vertices of a graph and put edges on a graph when a customer moves from $a_i$ to $a_j$ (i.e., $a_i \rightarrow a_j$). This graph is called a directed graph. Furthermore, it has cycles (i.e., $a_i \rightarrow a_i$) because it is possible to make more than one payment in the same area. By giving a weight, that is, the conditional probability of the location $a_j$ given $a_i$, on the edges of a graph, we can obtain an $N \times N$ square adjacency matrix. Fig. \ref{fig:F1} shows an example of a directed graph and the conditional probability of Table \ref{tab:T1} based on the adjacency matrix.

For instance, given the current transaction data in the region $\left \langle 6,4,1 \right \rangle$, the confidence score (or conditional probability) is $P(a=1|a=4)=0$, from Fig. \ref{fig:F1}.


\section{Model and algorithm}
\label{section:Sec4}
We compare the accuracy of the methods of AR model \citep{ref18} and GPs \citep{Gaussian06} for detecting large payments according to the transaction amount pattern, and also compare the accuracy of the methods of association rule analysis \citep{ref16} and adjacency matrix \citep{ref19} for detecting abnormal movement according to the transaction region pattern.

\begin{figure}[h!]
\begin{center}
\includegraphics[width=0.45\textwidth]{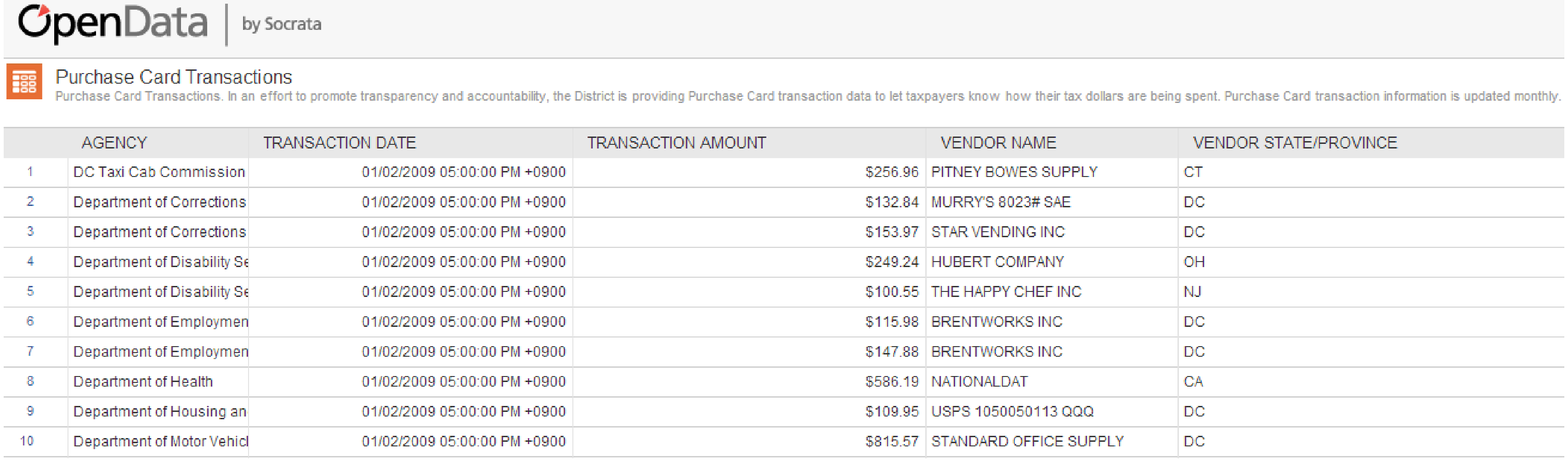}
\caption{OpenData used for the experiment}
\label{fig:F2}
\end{center}
\end{figure}

The data used for the experiments in this study were classified into two broad categories: legitimate transactions and fraudulent transactions. In our experiments, as legitimate transaction data, we used ``Purchase Card Transactions". We used the data in the ``Transaction amount" and ``Vendor State/Province" columns. The data are publicly available (Fig. \ref{fig:F2}, https://opendata.socrata.com/).

Fraudulent transaction data were extracted from the statistical results of \citet{ref3} using a dataset of real-world credit card transactions. The dataset contains 2,420 fraudulent transactions; a good summary of the dataset is given in \citet[Table 4]{ref3}. Given the fraudulent dataset, the transaction amount of fraudulent event is generated by non-negative distribution with the mean and standard deviation of the attribution, that is, the average amount spent per transaction over a month on all transactions up to this transaction. The transaction region of a fraudulent event is generated with the legitimate transaction location by random permutation based on the assumption that the transaction region pattern of perpetrators is distinguishable from that of a legitimate normal user of a card.

\begin{figure}[h!]
\begin{center}
\includegraphics[width=0.45\textwidth]{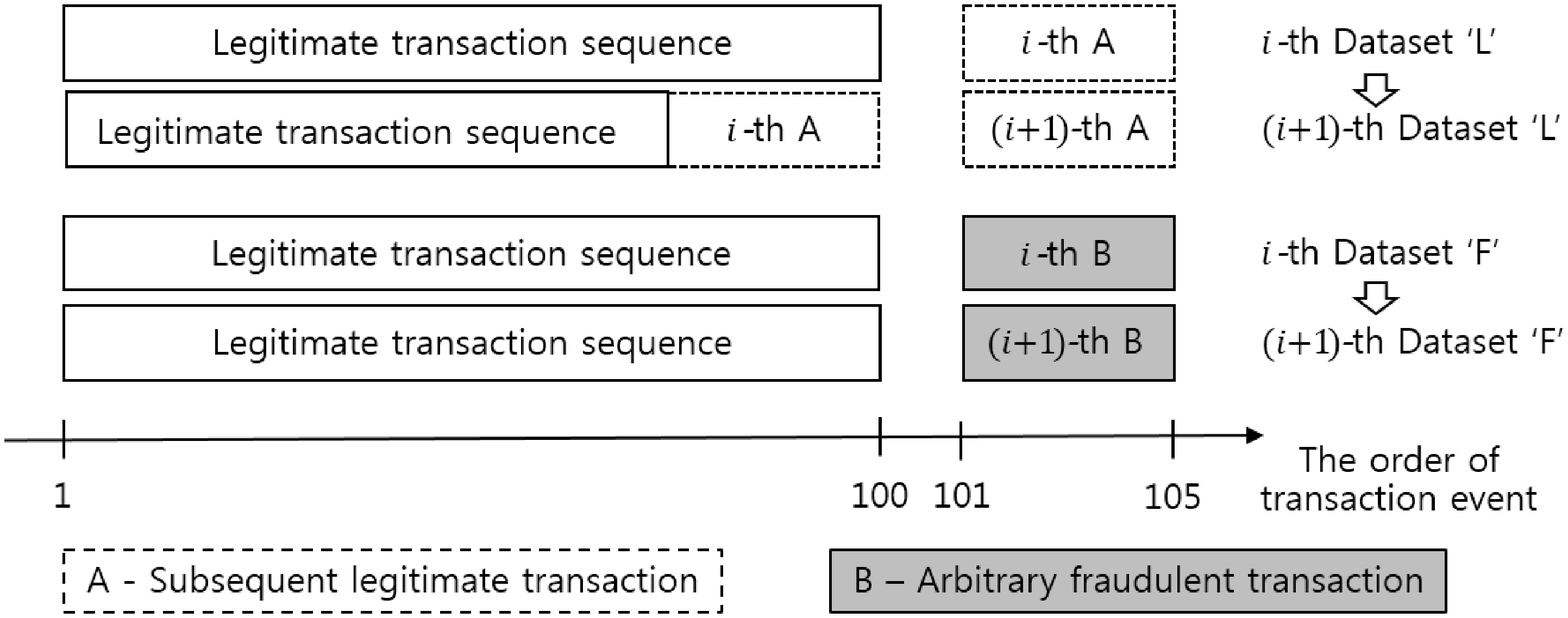}
\caption{Datasets used in experiments for detecting fraudulent transactions. Based on one legitimate transaction sequence, which consists of 100 transactions, we pad two kinds of data, subsequent legitimate transaction $A$ and arbitrary fraudulent transaction $B$, which consists of 5 transactions. We call them dataset $L$ and $F$ respectively. This figure also shows the process of constituting $(i+1)$-th dataset from $i$-th dataset. Each dataset $L$ preserves a certain transaction patterns by using a series of continuous legitimate transactions. In contrast, testing data of dataset $F$s has no relationship with training data (i.e., legitimate transaction sequence)}
\label{fig:F3}
\end{center}
\end{figure}

The datasets used our experiments are shown in Fig. \ref{fig:F3}. We first train with 100 transactions, which is called a ``legitimate transaction sequence", and learn the pattern of a legitimate card user. Then, we test the next transaction data whether it is legitimate or fraudulent. There are two kinds of testing data with length 5: ``subsequent legitimate transaction" data $A$ and ``arbitrary fraudulent transaction" data $B$. The legitimate transaction data $A$ are the subsequent continuous transaction data of the training data, and the fraudulent transaction data $B$ are independent of the previous training data. It should be noted that the number of the testing data is even smaller than that of the training data. Because perpetrators attempt to use stolen credit cards quickly to maximize the amount of the fraudulent payments, the sooner these transactions are detected, the greater the loss that can be avoided by stopping transactions made with the fraudulent credit cards \citep{ref3}.

Now, let the 105 legitimate transaction data be dataset $L$ and other transaction data inserted into the fraudulent transaction data be dataset $F$. We prepared 20 dataset $L$s and $F$s each to conduct our experiment for various transaction data. Both of dataset $L$s and $F$s include 105 transactions, and $(i+1)$-th dataset consists of 100 training data and 5 testing data as:
\begin{itemize}
\item $(i+1)$-th dataset $L =$ (rear part of legitimate transaction sequence $|$ 1-th $A$ $|$ $\cdots$ $|$ i-th $A$ $||$ $(i+1)$-th $A$)
\item $(i+1)$-th dataset $F =$ (original legitimate transaction sequence $||$ $(i+1)$-th $B$),
\end{itemize}
where ``$|$" represents a concatenation and ``$||$" represents a division between training data and testing data.

The characteristics of them are distinguishable with existence of transaction patterns. All dataset $L$s preserve a certain transaction patterns, but $F$s do not. While dataset $L$s consist of continuous legitimate transaction sequence, dataset $F$s consist of fraudulent transactions padded to a certain legitimate transaction sequence.

We conduct experiments using two methods to determine which method is better for detecting outliers that do not follow the pattern of the legitimate consumer and what threshold value is optimal for classifying the transactions of a legitimate customer and those of a perpetrator as:
\begin{itemize}
\item For the transaction amount, we set a threshold value and confirm whether the threshold allows a transaction amount to be identified as legitimate or fraudulent.
\item For transactions with amount and region, we give a confidence score for each transaction, that is, the similarity of the current transaction to previous transaction patterns, and confirm the distribution of these scores over legitimate and fraudulent cases. Then, we set a threshold based on the distribution.
\end{itemize}

In $AR(p)$ we should first make a decision over the number of AR terms, $p$, and the order of differencing, $d$. Here, $p$ represents the number of previous transactions used to express a transaction with a linear combination of previous transactions, and $d$ represents the number of differences needed to stationarize a sample of the legitimate transaction sequence.

The root mean square error (RMSE) shows the estimated white noise standard deviation, and the autocorrelation function (ACF) plot shows the coefficients of correlation between data and lags for the sample. It is helpful to decide $p$ and $d$ by focusing on the lowest standard deviation and the small and patternless autocorrelations.

\begin{table}[h]
\begin{center}
    \begin{tabular}{|c|c|}
    \hline
    $(p, d)$ & RMSE    \\ \hline
    (1, 1) & 1.92049 \\ 
    (2, 1) & 1.91002 \\ 
    (3, 0) & 1.64051 \\ 
    (4, 0) & 1.64892 \\ 
    \textbf{(5, 0)} & \textbf{1.61823} \\ \hline
    \end{tabular}
    \caption {The root mean square error (RMSE) for $p$ and $d$ over the sample of legitimate transaction sequence, where $p$ represents the number of AR terms and $d$ represents the order of differencing. The optimal value of $p$ and $d$ can be often founded at which the lowest RMSE.}
    \label{tab:rmse}
\end{center}
\end{table}

\begin{figure}[h!]
\centering
\includegraphics[width=0.45\textwidth]{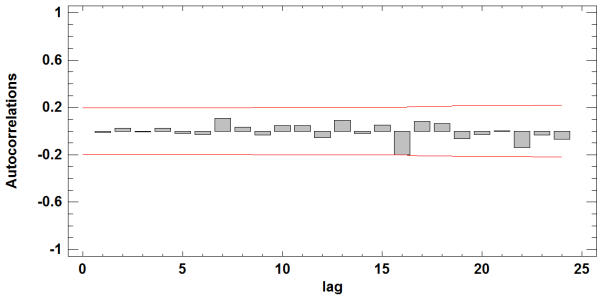}
\caption{The bar graph of autocorrelations of AR model with $p=5$ and $d=0$ for transaction amount of the sample. Two red horizontal lines indicate the approximate upper and lower confidence bounds. If the autocorrelations are all small and patternless, then the data does not need a higher order of differencing.}
\label{fig:autocorrelation}
\end{figure}

According to Table \ref{tab:rmse} and Fig. \ref{fig:autocorrelation}, we choose $p=5$ and $d=0$.

In the AR model and GPs, we calculate the estimated mean and variance of $y_{N+1}$, $E$ and $V$ respectively, EVP of $y_{N+1}$, $P$ and the confidence score of $y_{N+1}$, $C$ from the training data, $Y=\left \{ y_i \right \}_{i=1}^N$, and testing data, $y_{N+1}$ using algorithm 1 and 2.
\begin{algorithm}[h!]
Autoregressive() \\
$p \leftarrow 5$ and $Y \gets \ln Y$\\
$X \leftarrow \left \{ 1, y_i \right \}_{i=1}^N$ as $X$ of equation \eqref{eq:linear} \\
$\hat{A} \leftarrow$ equation \eqref{eq:hat_A} and $\tau^{-1} \leftarrow$ equation \eqref{eq:tau} \\
$E \leftarrow X\hat{A}$ \\
$V \leftarrow \tau^{-1}$ \\
$P \leftarrow P_{EV}(y_{N+1})$ as equation \eqref{eq:EVP} \\
$C \leftarrow 1-cdf(\text{Normal},y_{N+1}, E,V)$ \\
return $E$, $V$, $P$, $C$
\caption{Predict and score the confidence of transaction amount using the AR model}
\end{algorithm}

In algorithm 1, ``$cdf(\text{Normal},y_{N+1}, E,V)$" represents a cumulative distribution function value of normal distribution with mean $E$ and variance $V$ at $y_{N+1}$.

\begin{algorithm}[h]
GaussianProcess() \\
$\mathbf{x}\leftarrow \left ( 1\: 2\: \cdots N \right )^T$ \\
$\mathbf{y}\leftarrow \ln Y$ \\
$K \leftarrow$ equation \eqref{eq:22} \\
$l,\, \sigma_f,\, \sigma_n \leftarrow \arg\max\left \{ \ln p(\theta|\mathbf{x},\mathbf{y})\right \}$ as equation \eqref{eq:para_theta}\\
$k(x,x')\leftarrow$ equation \eqref{eq:20} \\
$K,\, K_*,\, K_{**} \leftarrow$ equation \eqref{eq:22}, \eqref{eq:23}, and \eqref{eq:24} \\
$E \leftarrow K_*K^{-1}\mathbf{y}$ \\
$V \leftarrow K_{**}-K_*K^{-1}K_*^T$ \\
$P \leftarrow P_{EV}(y_{N+1})$ as equation \eqref{eq:EVP} \\
$C \leftarrow 1-cdf(\text{Normal},y_{N+1}, E,V)$ \\
return $E$, $V$, $P$, $C$
\caption{Predict and score the confidence of transaction amount using GPs}
\end{algorithm}

In the association rule and adjacency matrix, we calculate only the confidence score $C$ of the transaction region $a_{N+1}$ of the testing data from transaction path $\left \{ a_i \right \}_{i=1}^{N}$ of the training data using algorithms 3 and 4. In algorithm 3, $\mathbf{U}_i$ means the $i$-th row of matrix $\mathbf{U}$. In algorithm 4, $A(i,j)$ means an entry value of the $i$-th row, $j$-th column of matrix $A$.

\begin{algorithm}[h!]
AssociationRule() \\
\For{($i=1 \to N$)}
{
$\mathbf{U} \leftarrow a_i$ with 10 columns as equation \eqref{eq:33}
}
\ForEach{subsequence $s$ of $\left \langle a_{N-1}, a_{N}, a_{N+1} \right \rangle$}
{
	$i \gets 1$ \\
	\While{$\mathbf{U}_i\neq \varnothing$}
	{
		\Case{$s\subset \mathbf{U}_i$}
			{
				$S_{\left \langle s \right \rangle}\gets S_{\left \langle s \right \rangle}+1$
			}
		\Case{for $a\in s$, $a\in \mathbf{U}_i$}
			{
			$S_{\left \langle s \right \rangle}\gets S_{\left \langle s \right \rangle}+\frac{1}{1+t}$	as equation \eqref{eq:36}
			}
	$i \gets i+1$
	}
}
$C_1 \gets S_{\left \langle a_{N-1}, a_N, a_{N+1} \right \rangle}/S_{\left \langle a_{N-1} \right \rangle}$ \\
$C_2 \gets S_{\left \langle a_{N-1}, a_N, a_{N+1} \right \rangle}/S_{\left \langle a_{N-1}, a_N \right \rangle}$ \\
$C_3 \gets S_{\left \langle a_N, a_{N+1} \right \rangle}/S_{\left \langle a_N \right \rangle}$ \\
$C \gets \max \left \{ C_1,C_2,C_3 \right \}$ \\
return $C$
\caption{Score the confidence of transaction region using association rule}
\end{algorithm}
\begin{algorithm}[h]
AdjacencyMatrix() \\
$A \gets N\times N$ zero matrix \\
\For{$i=1 \to N$}
{
	$A(a_i,a_{i+1})\gets A(a_i,a_{i+1})+1$
}
\ForEach{$A(i,j)$}
{
	$A(i,j) \gets A(i,j)/\sum_{j=1}^{N}A(i,j)$
}
$C \gets A(a_N, a_{N+1})$ \\
return $C$
\caption{Score the confidence of transaction region using adjacency matrix}
\end{algorithm}

\section{Results}
\label{section:Sec5}

\subsection{Outlier detection for the transaction amount}

\begin{figure}[h!]
\begin{tabular}{cc}
\includegraphics[scale=0.235]{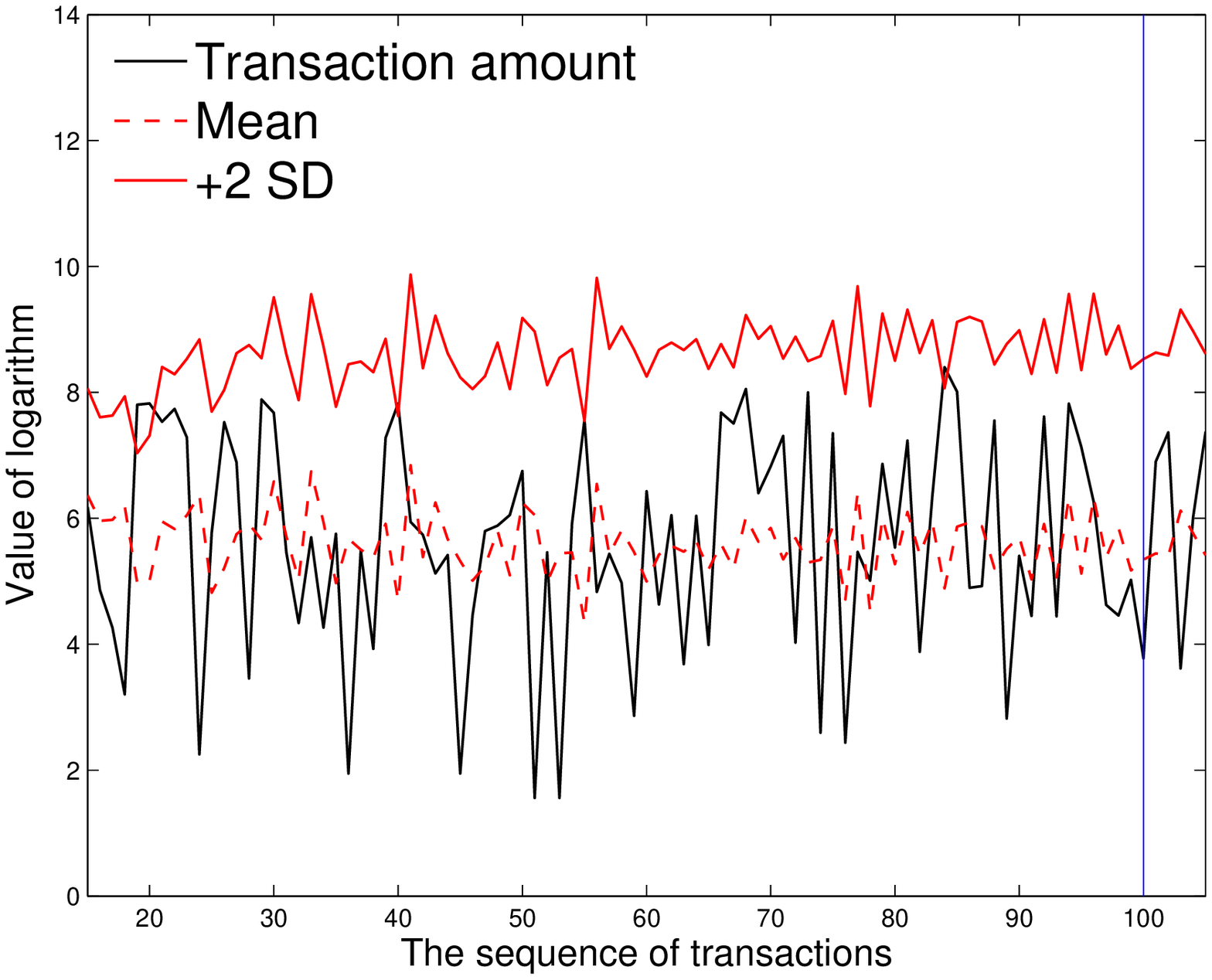} & 
\includegraphics[scale=0.235]{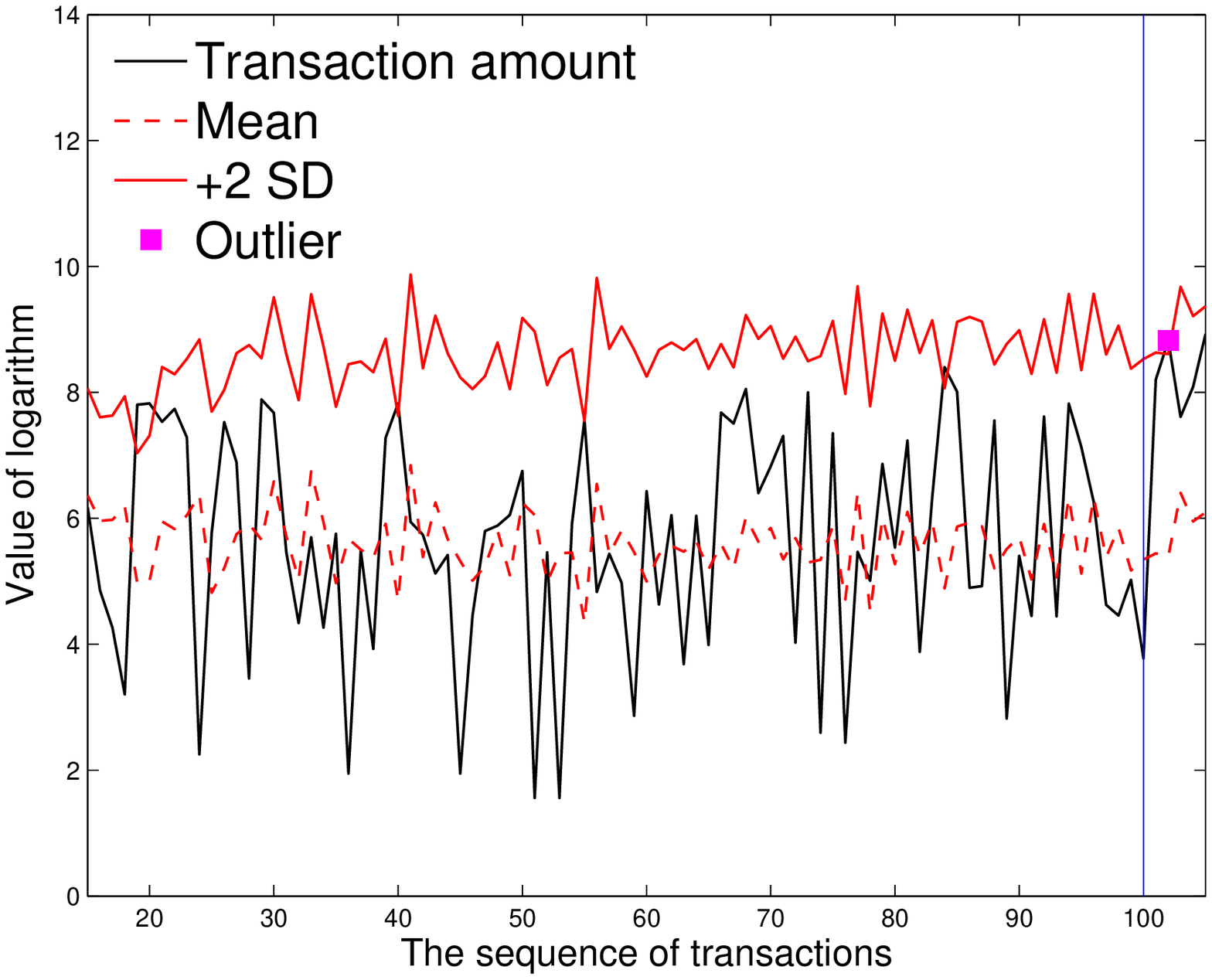} \cr
\includegraphics[scale=0.235]{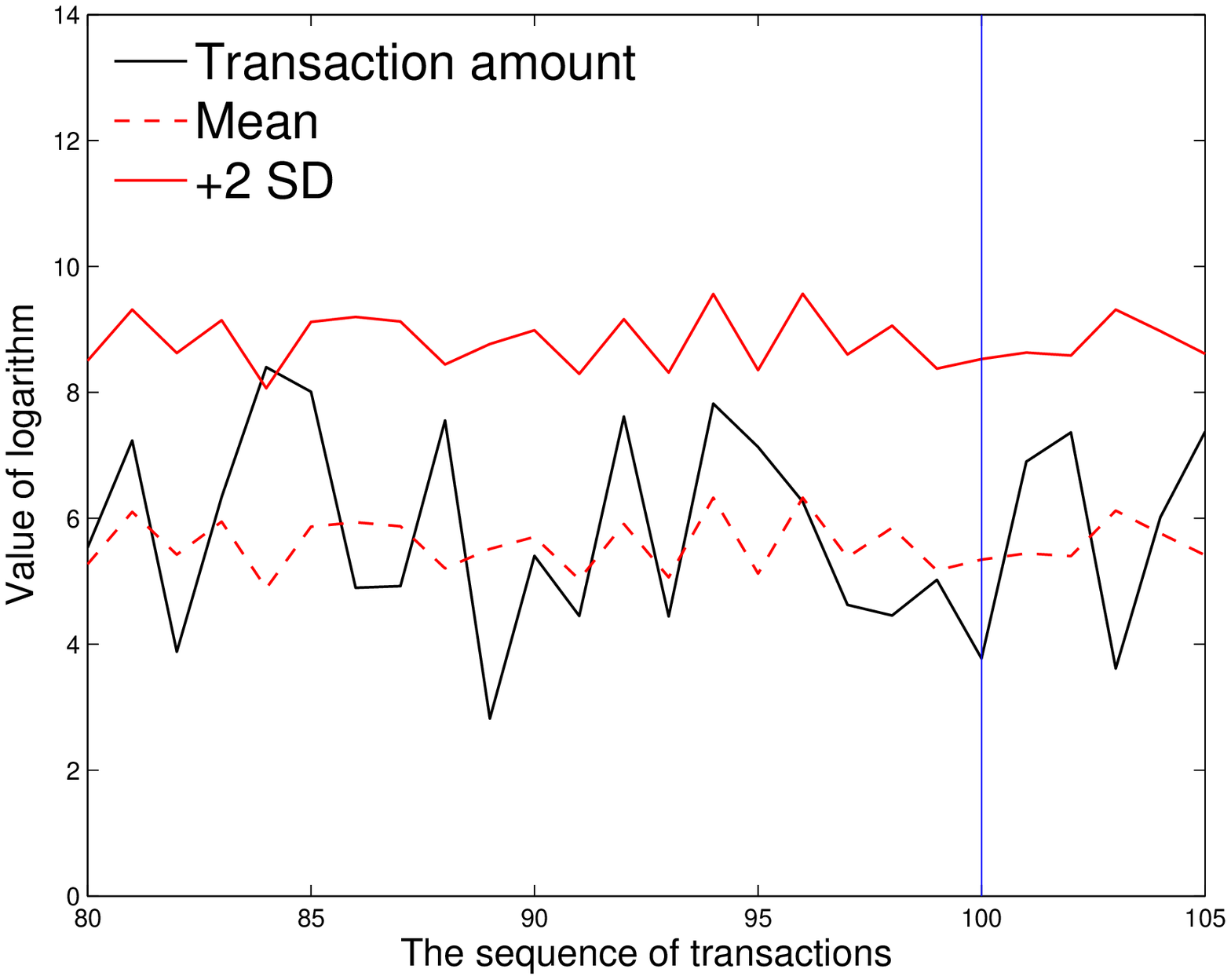} & 
\includegraphics[scale=0.235]{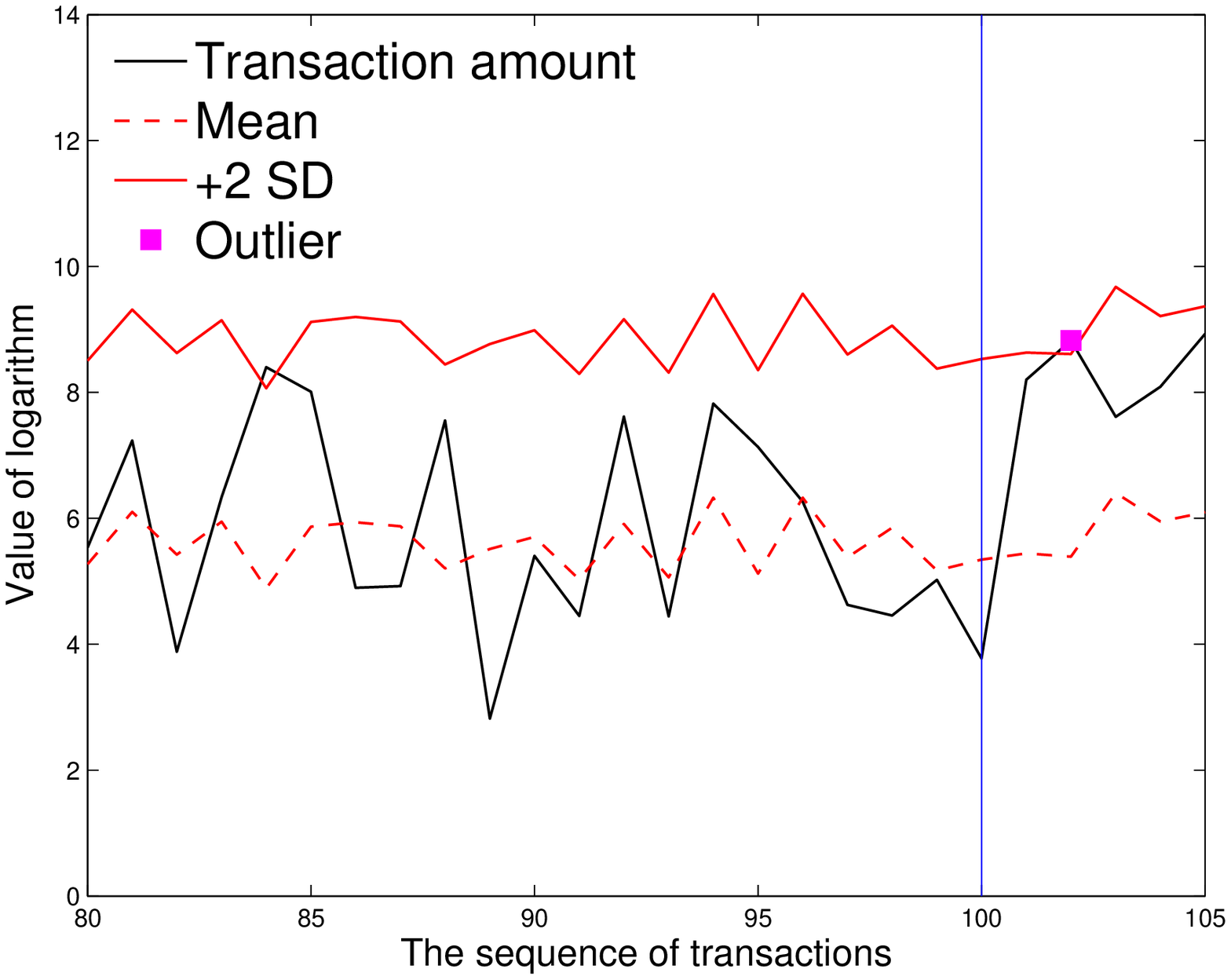} \cr
(a) Dataset $L$ &
(b) Dataset $F$
\end{tabular}
\caption{Outlier detection with $AR(5)$ and +2 SD. The transaction amount of datasets $L$ and $F$ are represented by the solid black line. The vertical blue line divides the transactions into the legitimate transaction sequence on the left side and subsequent legitimate transactions on the right side in (a), or arbitrary fraudulent transactions on the right side in (b). The mean and 2 standard deviation estimated from $AR(5)$ are represented by the red dotted and solid line respectively.}
\label{fig:result1}
\end{figure}

\begin{figure}[h!]
\begin{tabular}{cc}
\includegraphics[scale=0.235]{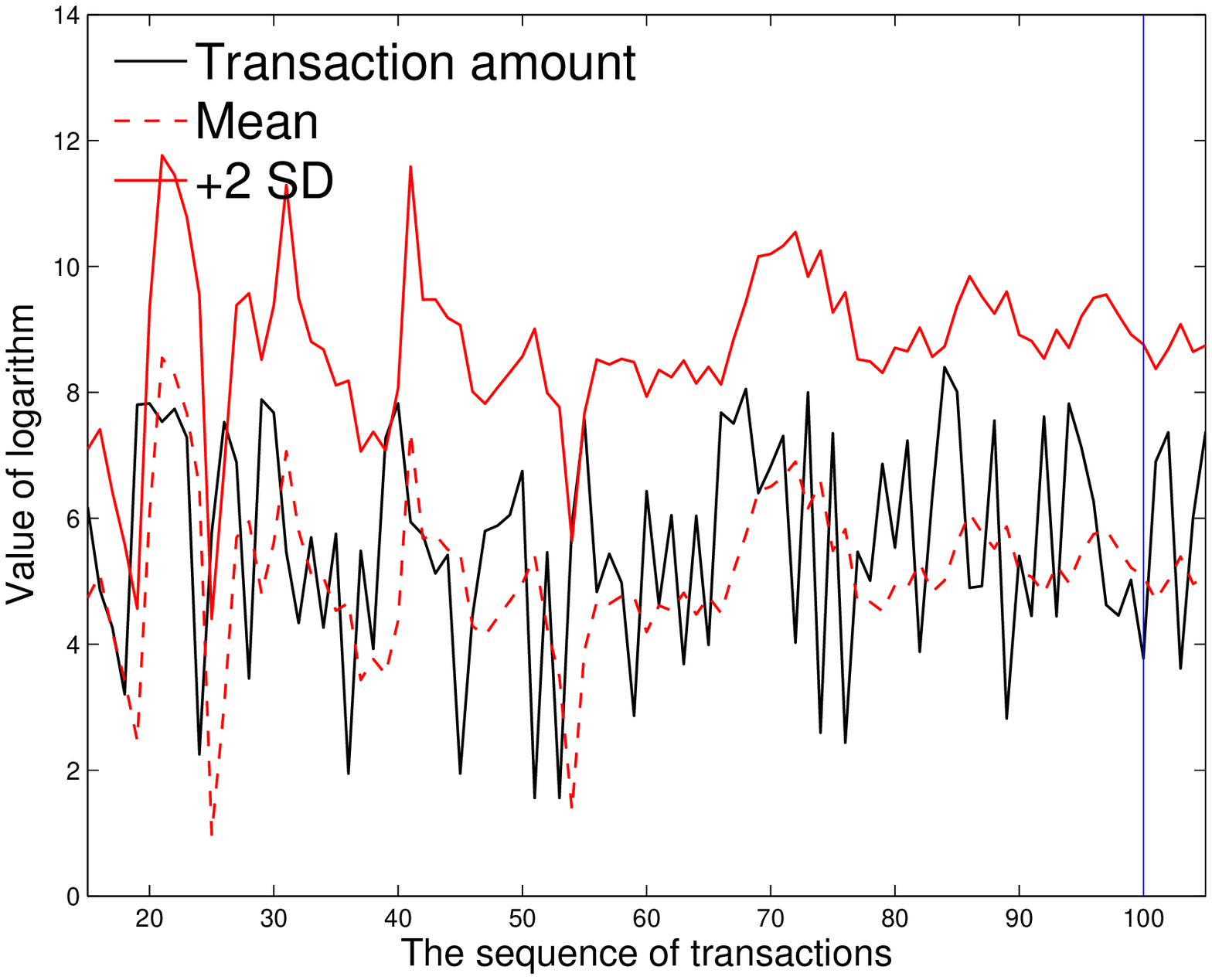} & 
\includegraphics[scale=0.235]{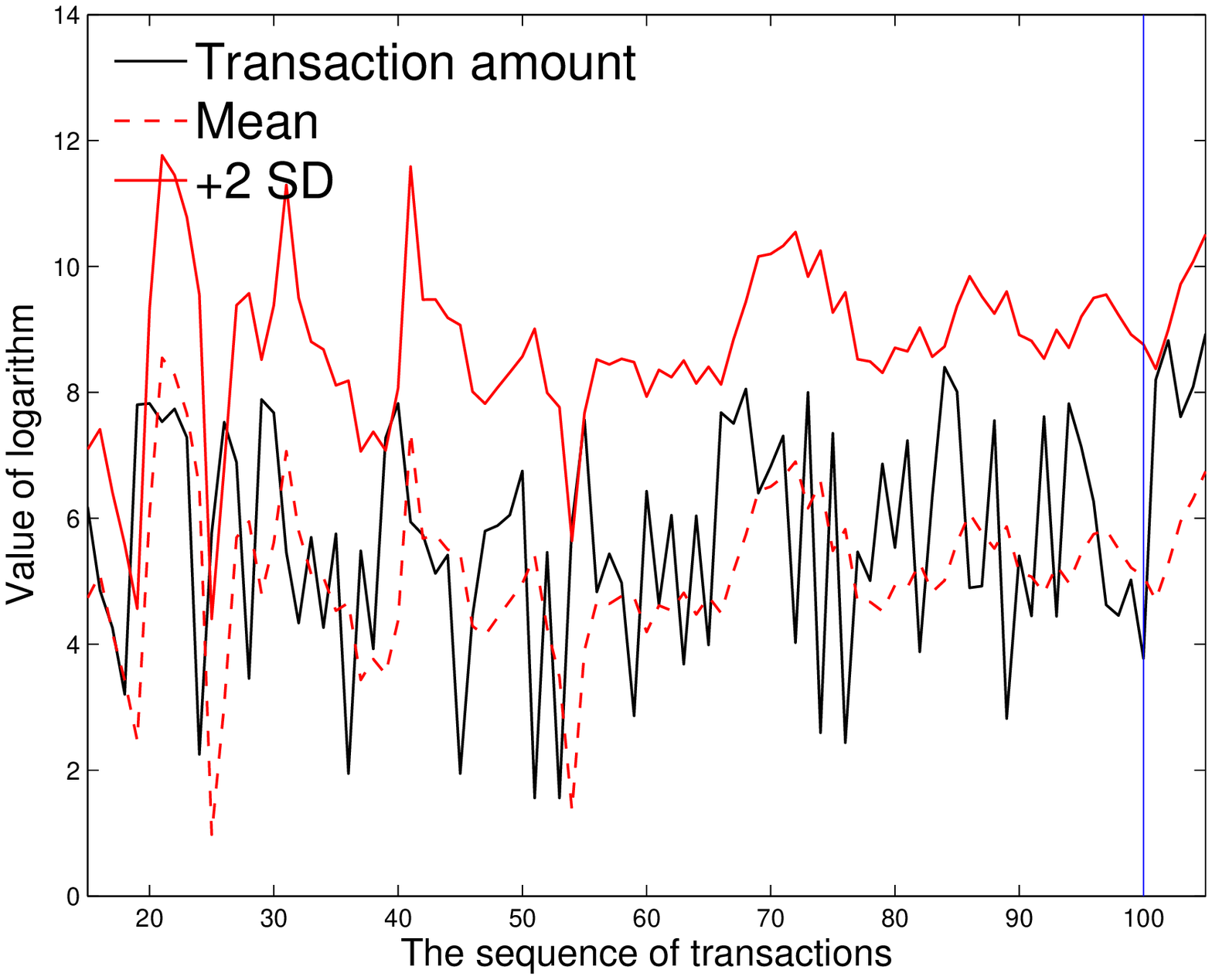} \cr
\includegraphics[scale=0.235]{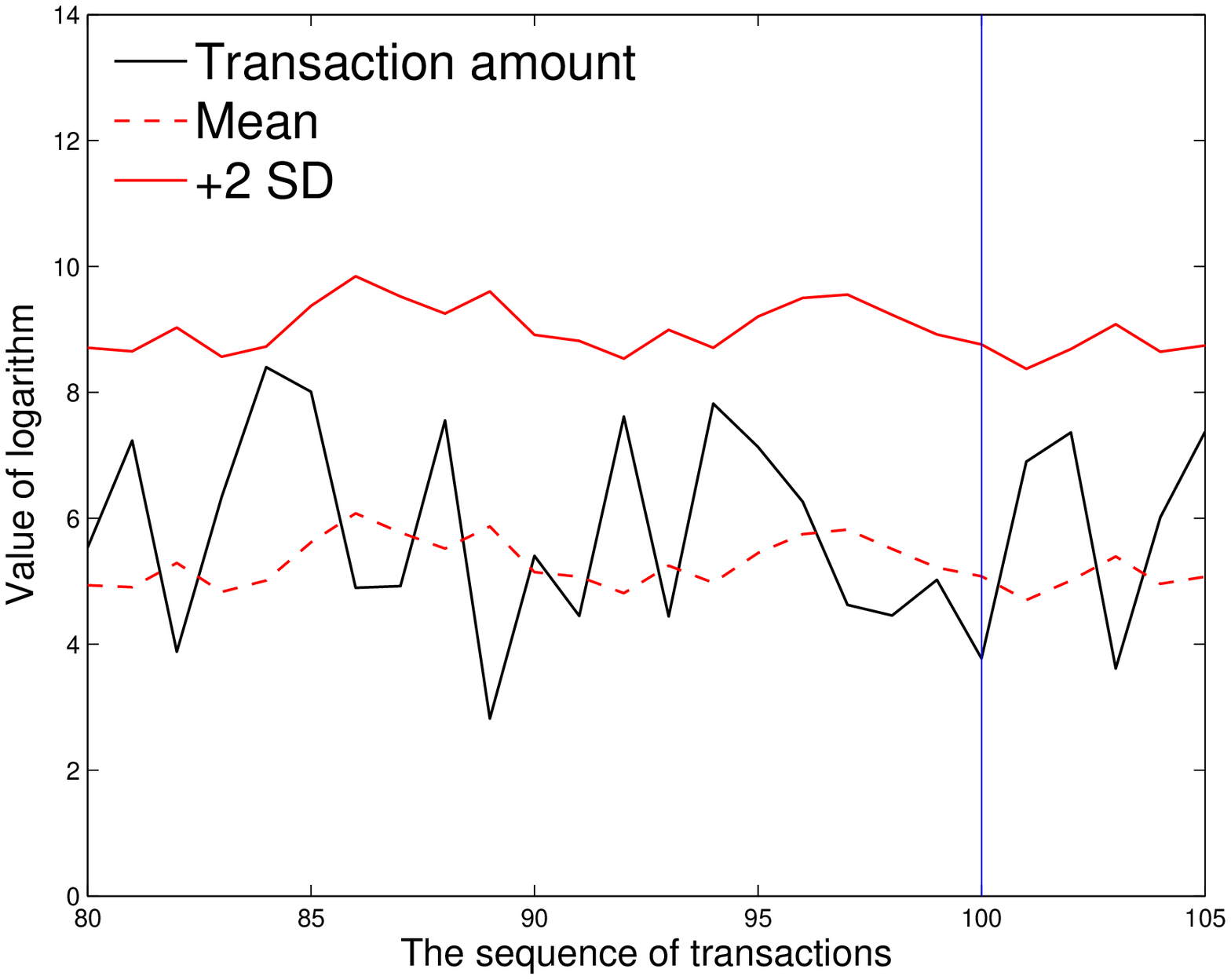} & 
\includegraphics[scale=0.235]{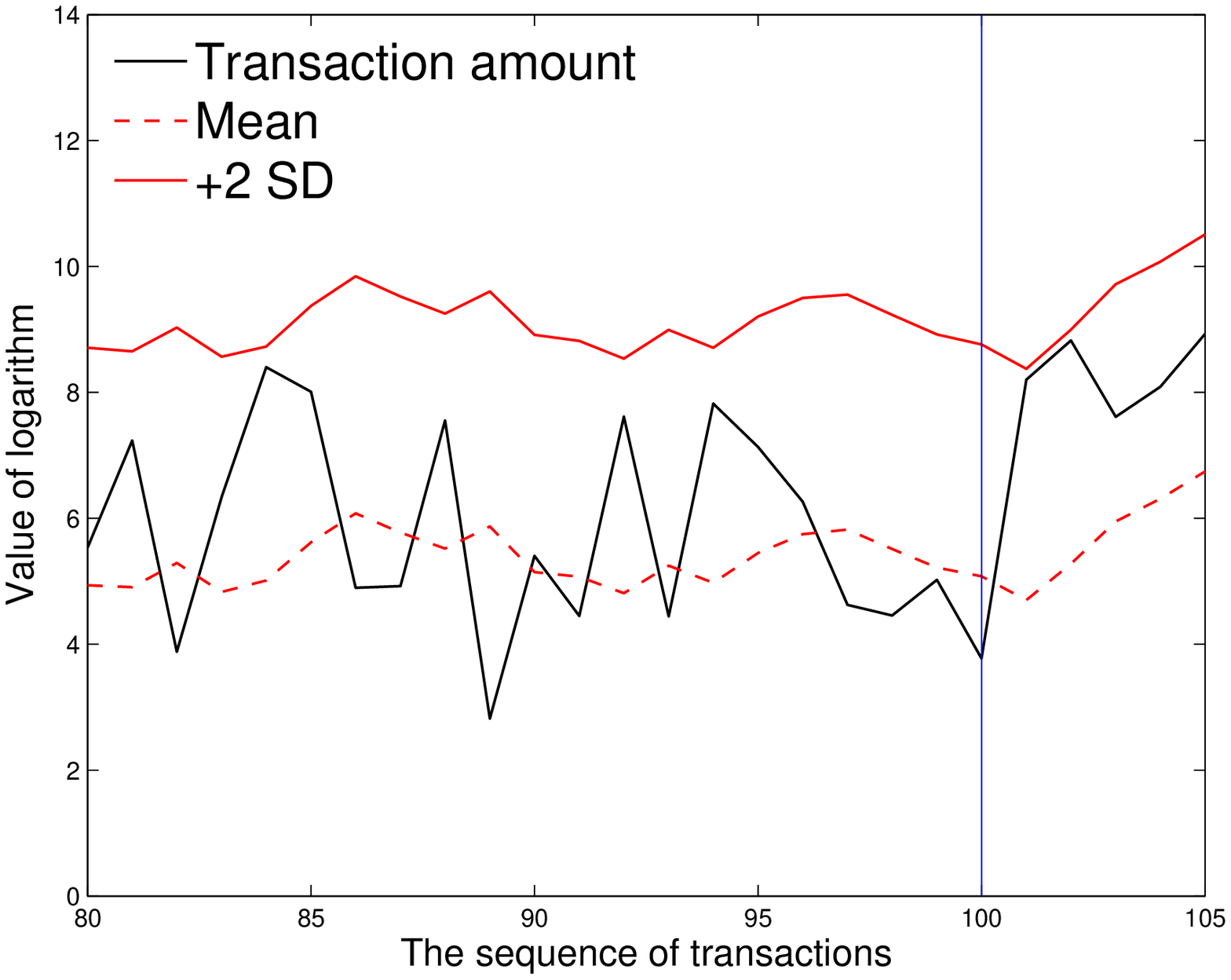} \cr
(a) Dataset $L$ &
(b) Dataset $F$
\end{tabular}
\caption{Outlier detection with GPs and +2 SD. The red dotted and solid lines represent, respectively, the mean and 2 standard deviation estimated from GPs. The datasets $L$ and $F$ are same as those used in $AR(5)$.}
\label{fig:result2}
\end{figure}

We first apply the logarithmic functions to datasets $L$ and $F$ in the experiment, and perform outlier detection for the transaction amount using GPs and $AR(5)$ at a confidence level of 95\% (i.e., +2 SD), as shown in Fig. \ref{fig:result1} and \ref{fig:result2}. The upper plots in these figures help us to obtain a comprehensive grasp of the transaction amount and its mean and upper bound, while the lower plots focus on the range of testing data, $A$ and $B$, from the 101-st to 105-th transaction.

In the case of $AR(5)$, as shown in Fig. \ref{fig:result1}, all the legitimate transactions in dataset $L$ are identified as legitimate. However, in dataset $F$, the fraudulent transactions are not detected as such except one transaction, because the upper bound for dataset $F$ was too high to identify the amount of the fraudulent transactions.

In the case of GPs, as shown in Fig. \ref{fig:result2}, GPs also identified the legitimate transactions in dataset $L$, and not the fraudulent transactions in dataset $F$. It seems that the arbitrary fraudulent transactions, $B$, in dataset $F$ do not overwhelm the large amount of legitimate transaction sequences that sometimes occurred.

\begin{figure}[h!]
\begin{tabular}{cc}
\includegraphics[scale=0.235]{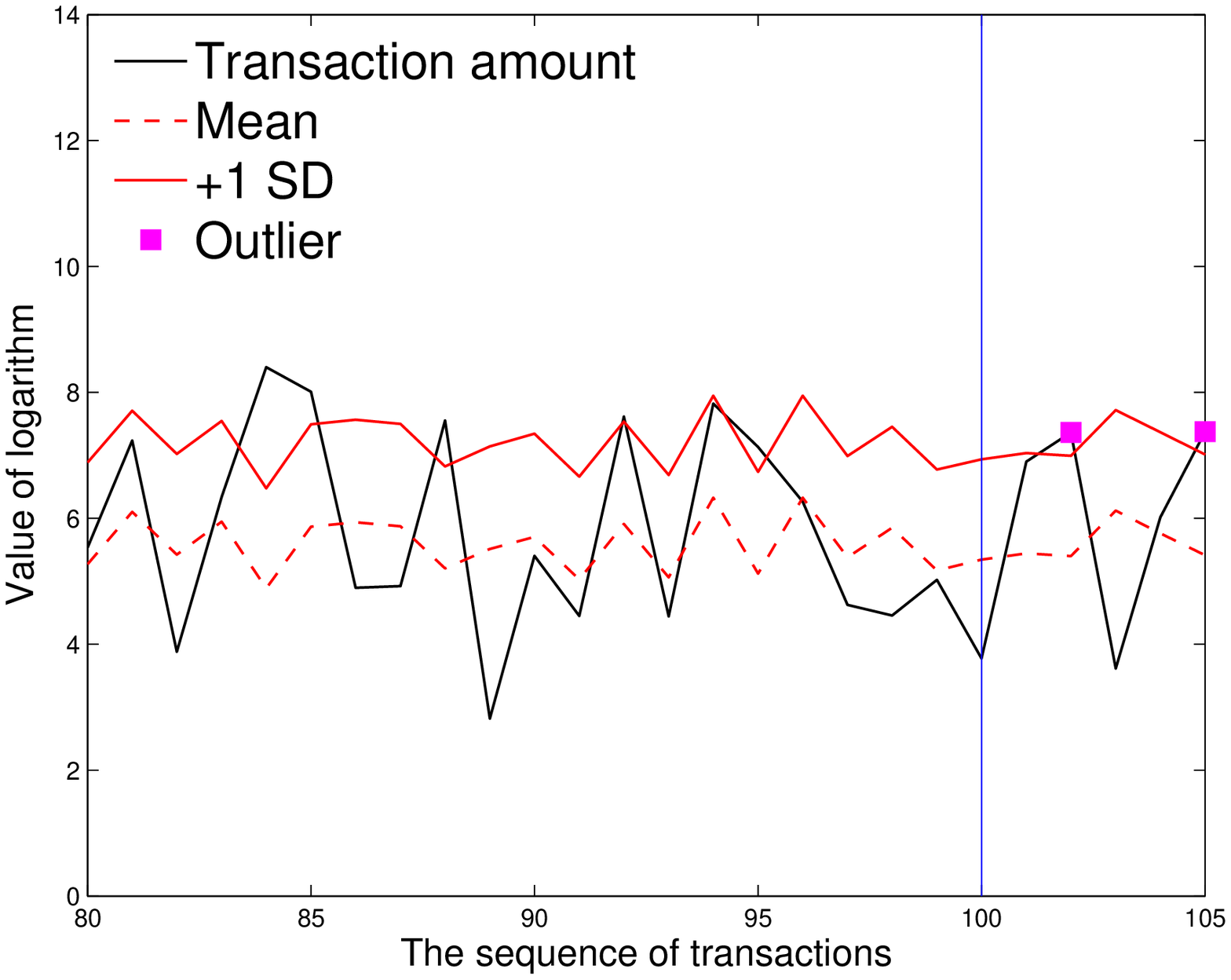} & 
\includegraphics[scale=0.235]{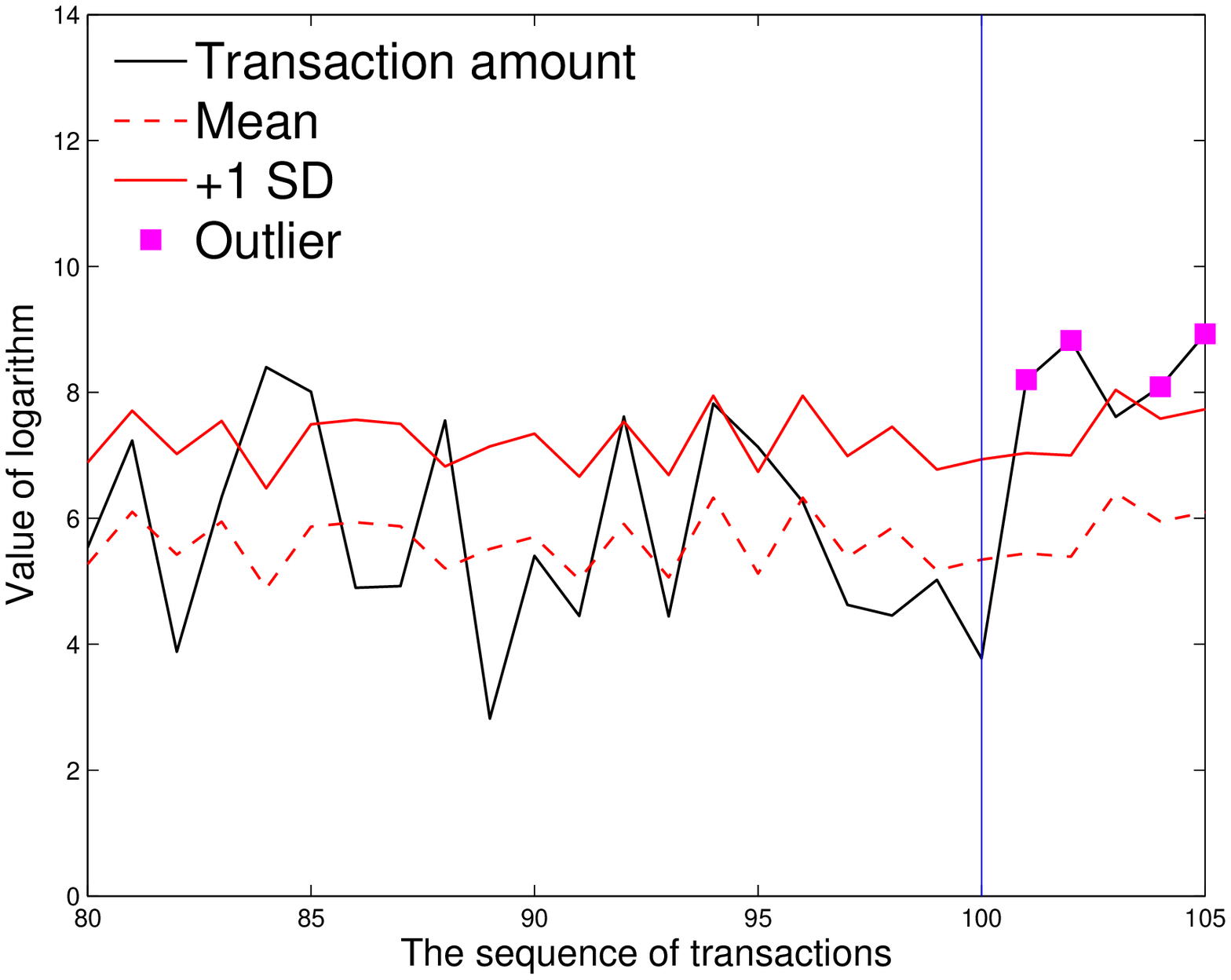} \cr
(a) Dataset $L$ &
(b) Dataset $F$
\end{tabular}
\caption{Outlier detection with $AR(5)$ and +1 SD. Transaction amount and estimated mean are same as in Fig.\ref{fig:result1}. Only the upper bound is downsized.}
\label{fig:result3}
\end{figure}

\begin{figure}[h!]
\begin{tabular}{cc}
\includegraphics[scale=0.235]{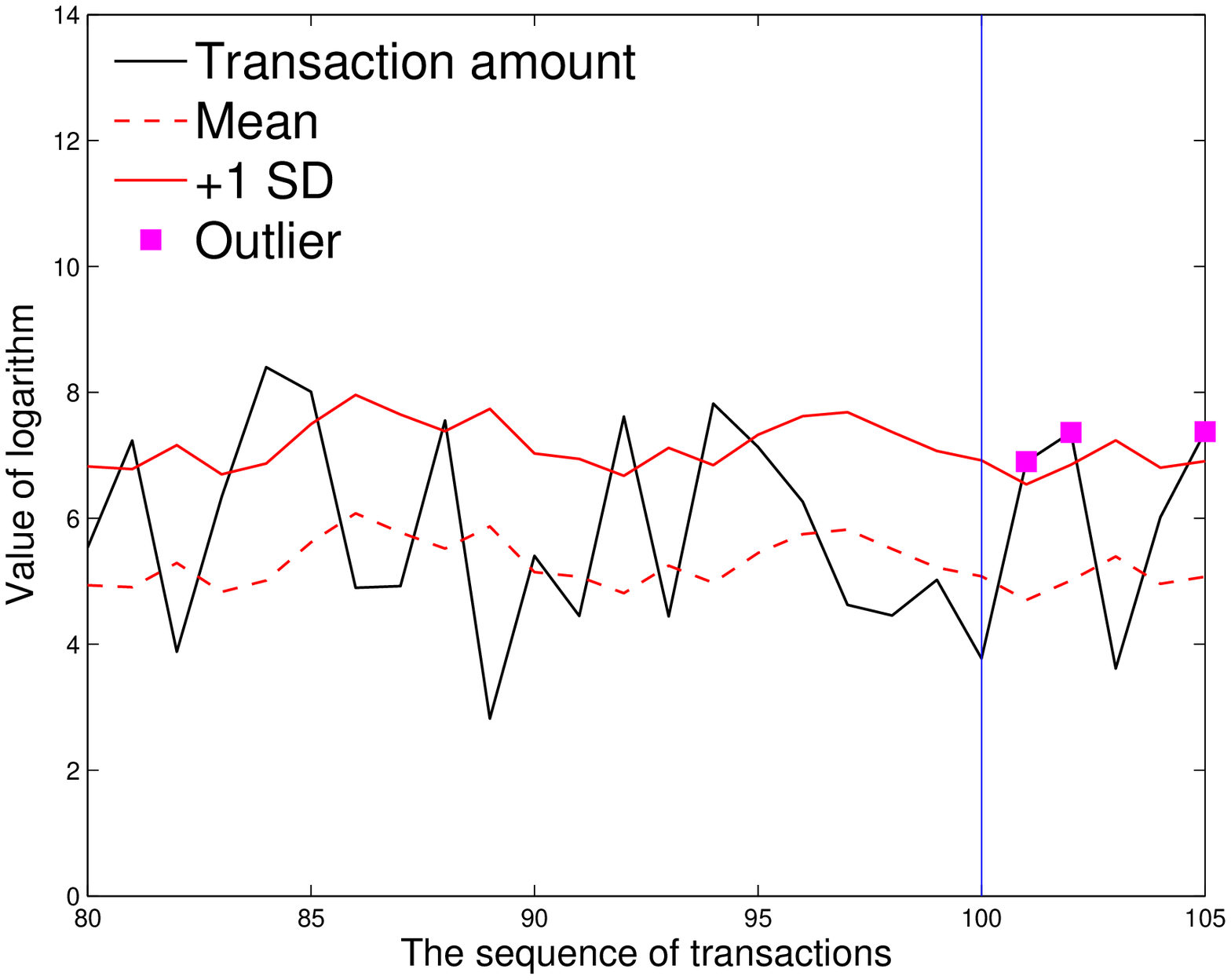} & 
\includegraphics[scale=0.235]{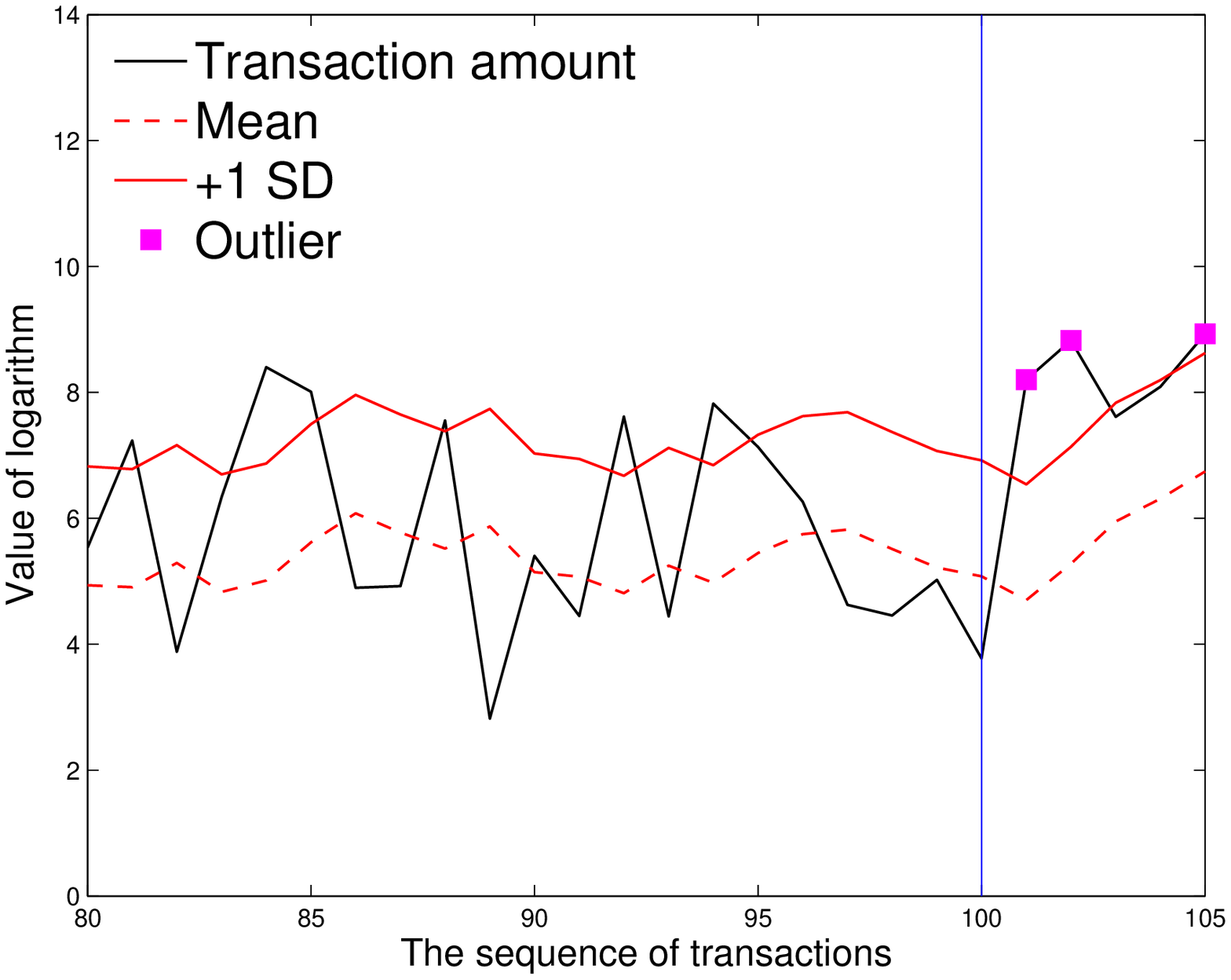} \cr
(a) Dataset $L$ &
(b) Dataset $F$
\end{tabular}
\caption{Outlier detection with GPs and +1 SD. Transaction amount and estimated mean are same as in Fig. \ref{fig:result2}. Only the upper bound is downsized.}
\label{fig:result4}
\end{figure}

When the confidence level threshold is adjusted to 68\% (i.e., +1 SD), in Figs. \ref{fig:result3} and \ref{fig:result4}, outliers, indicated by magenta square, are found more easily than when threshold is 95\% (i.e., +2 SD). However, false-positive errors were incurred because the legitimate transactions were detected, as well as the dataset $F$ that contains fraudulent transactions.

\begin{figure}[h!]
\begin{tabular}{cc}
\includegraphics[scale=0.235]{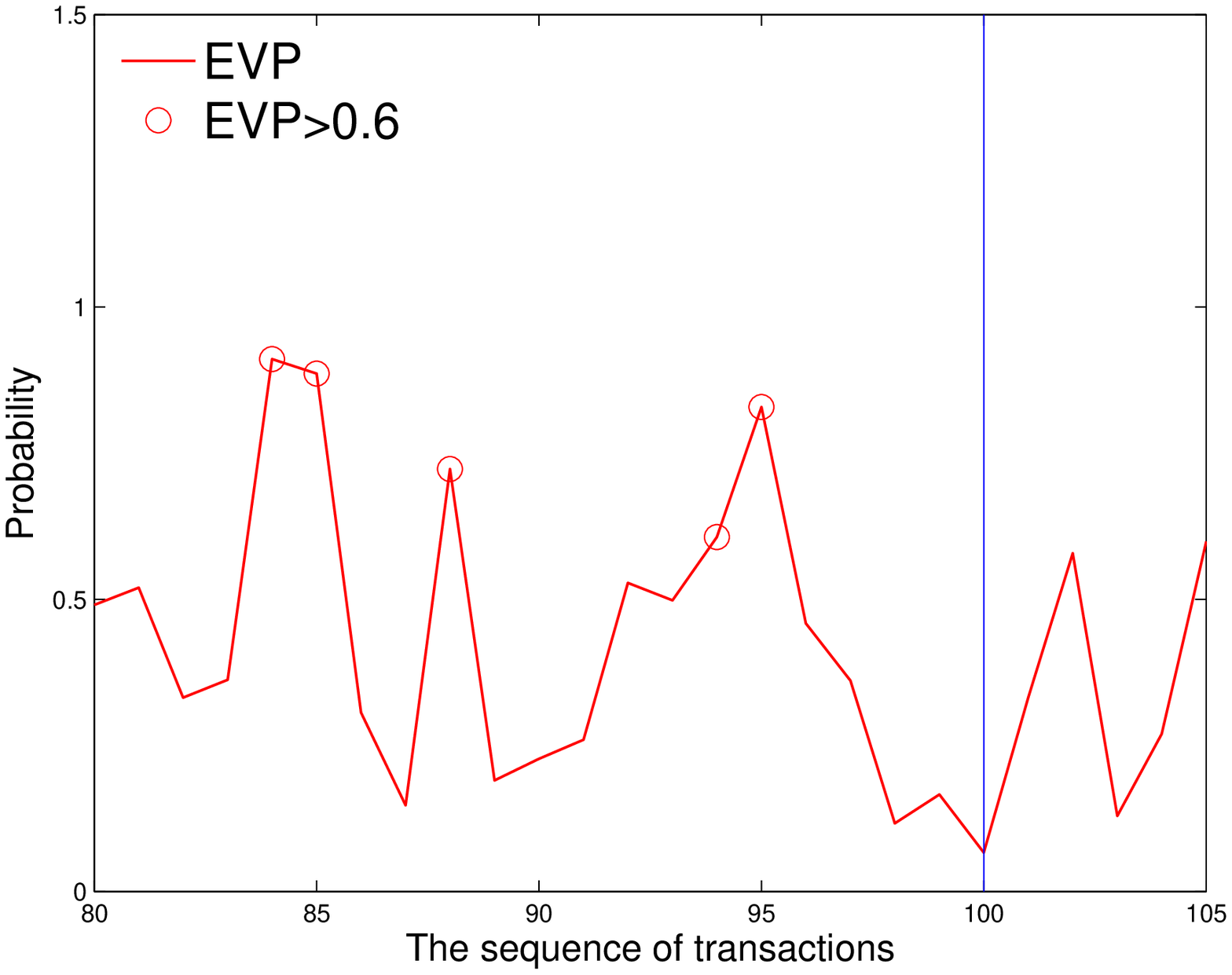} & 
\includegraphics[scale=0.235]{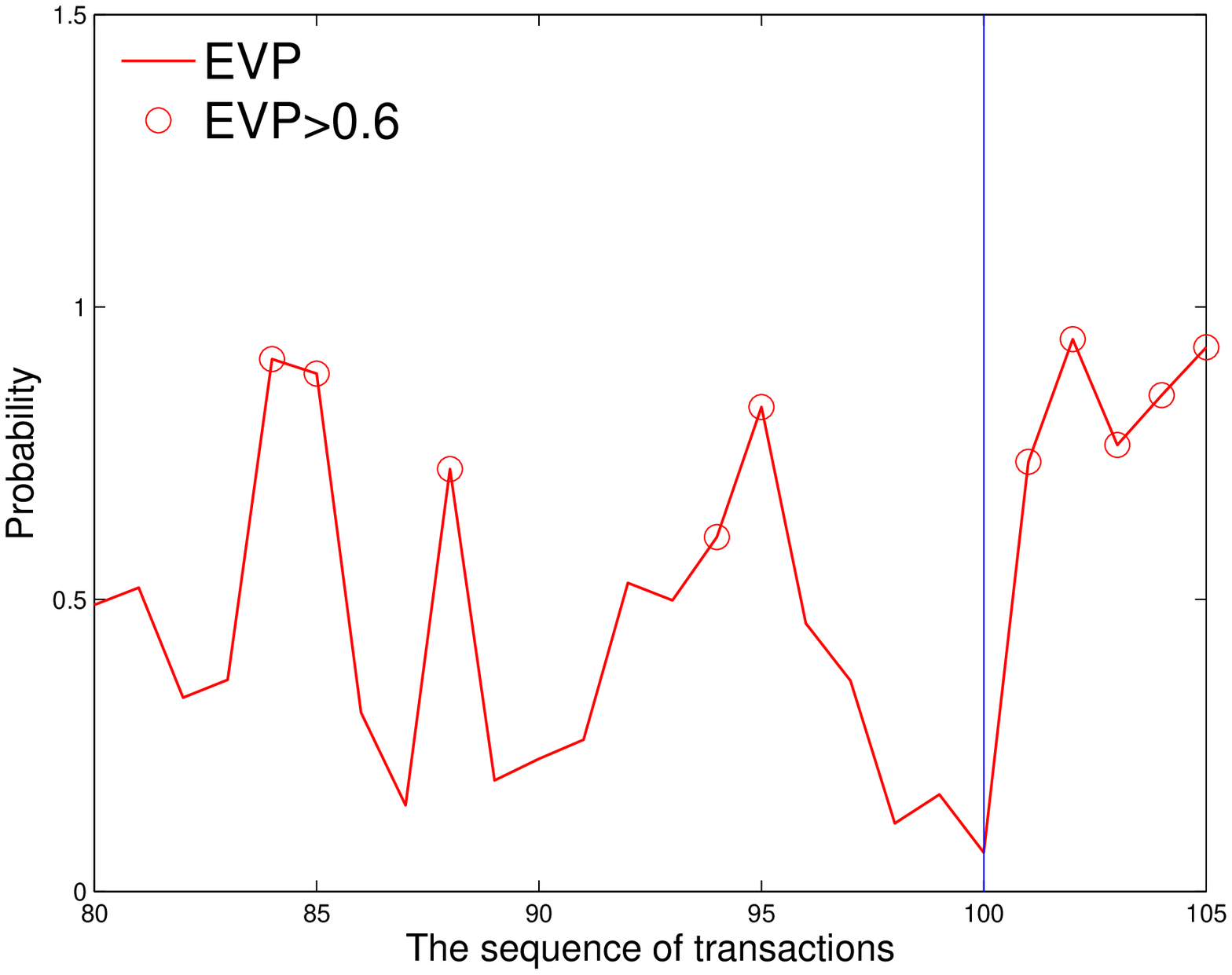} \cr
\includegraphics[scale=0.235]{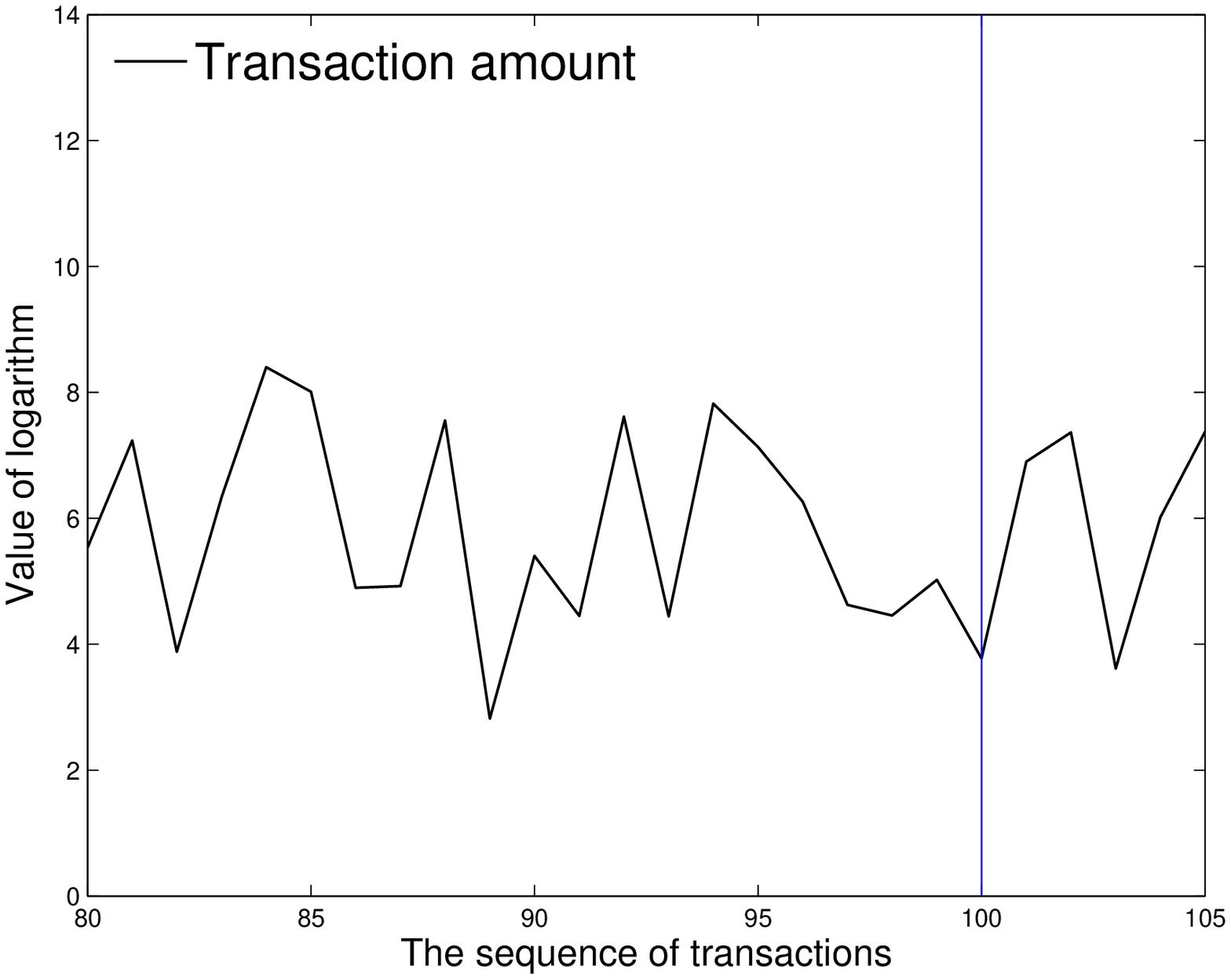} & 
\includegraphics[scale=0.235]{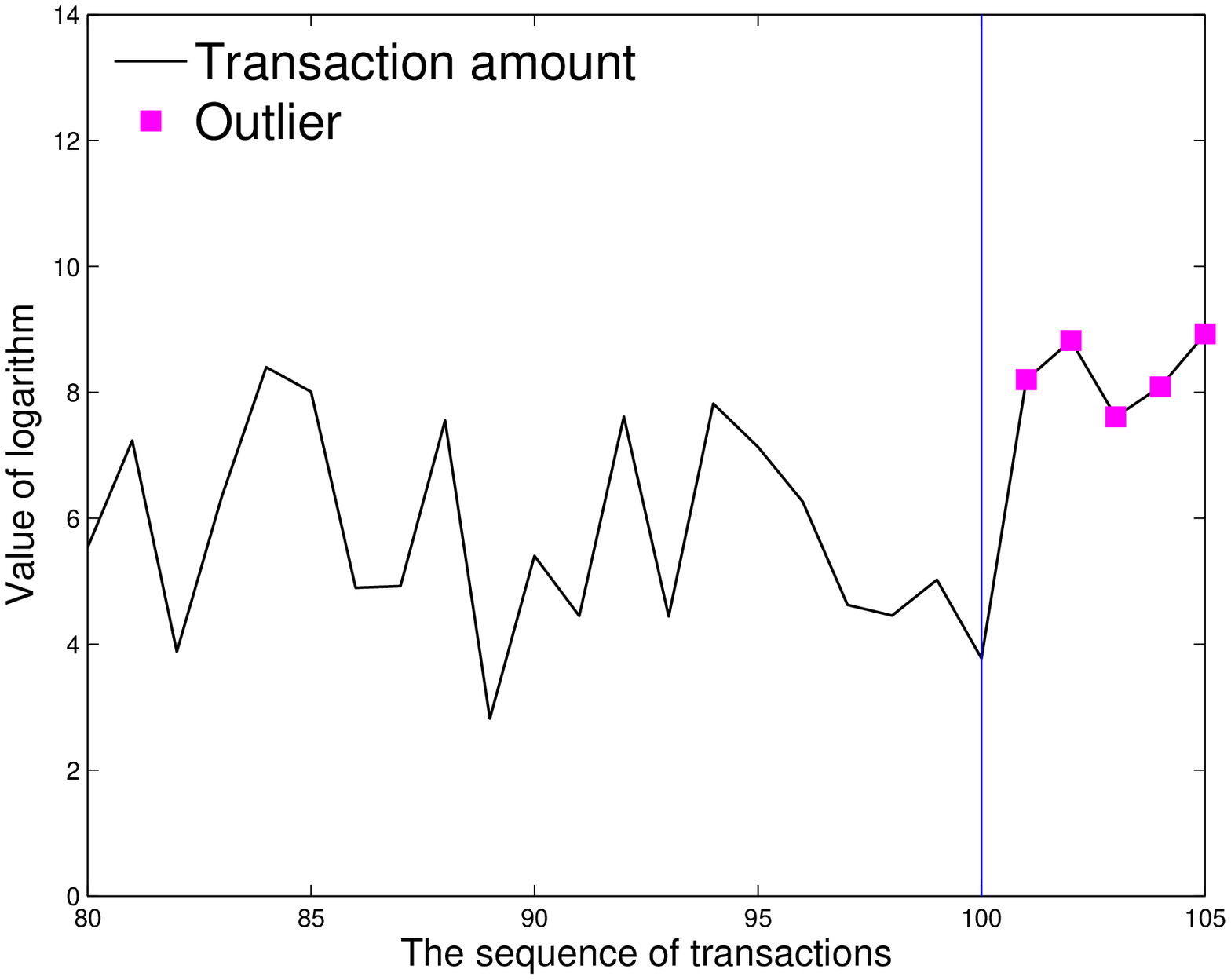} \cr
(a) Dataset $L$ &
(b) Dataset $F$
\end{tabular}
\caption{Outlier detection with $AR(5)$ and EVP. The upper plots show the extreme value probability of the transaction amount represented by the solid red line. Means and standard deviations of each transaction amount are used to draw EVP and obtained by $AR(5)$. Red circles indicate the points where $P_{EV}$ is larger than 0.6. The lower transactions corresponding to these points are defined as outliers. They are indicated by magenta squares in the lower plot. The lower plots show the transaction amounts and outliers, which are transactions where $P_{EV}$ is larger than 0.6.}
\label{fig:result5}
\end{figure}

\begin{figure}[h!]
\begin{tabular}{cc}
\includegraphics[scale=0.235]{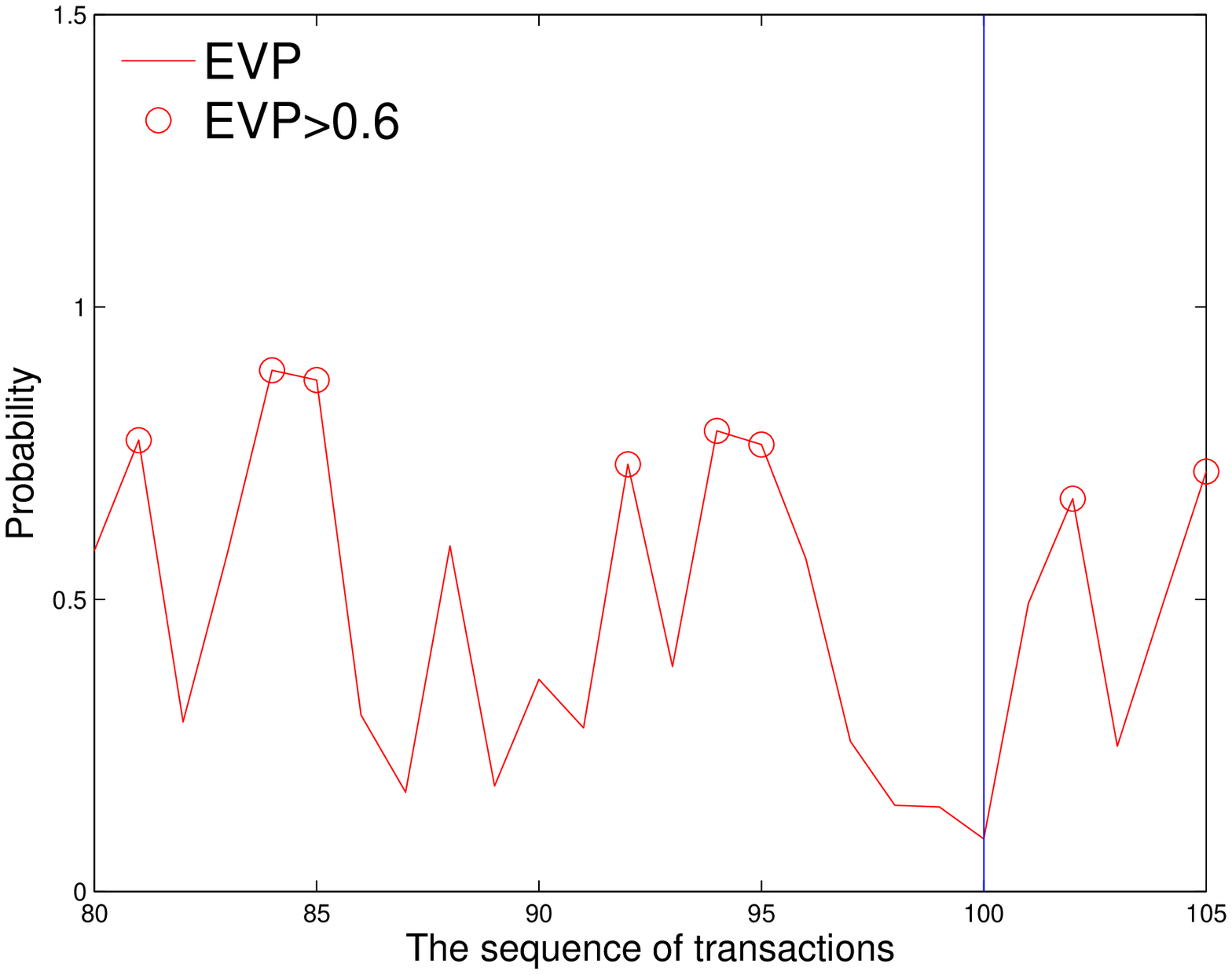} & 
\includegraphics[scale=0.235]{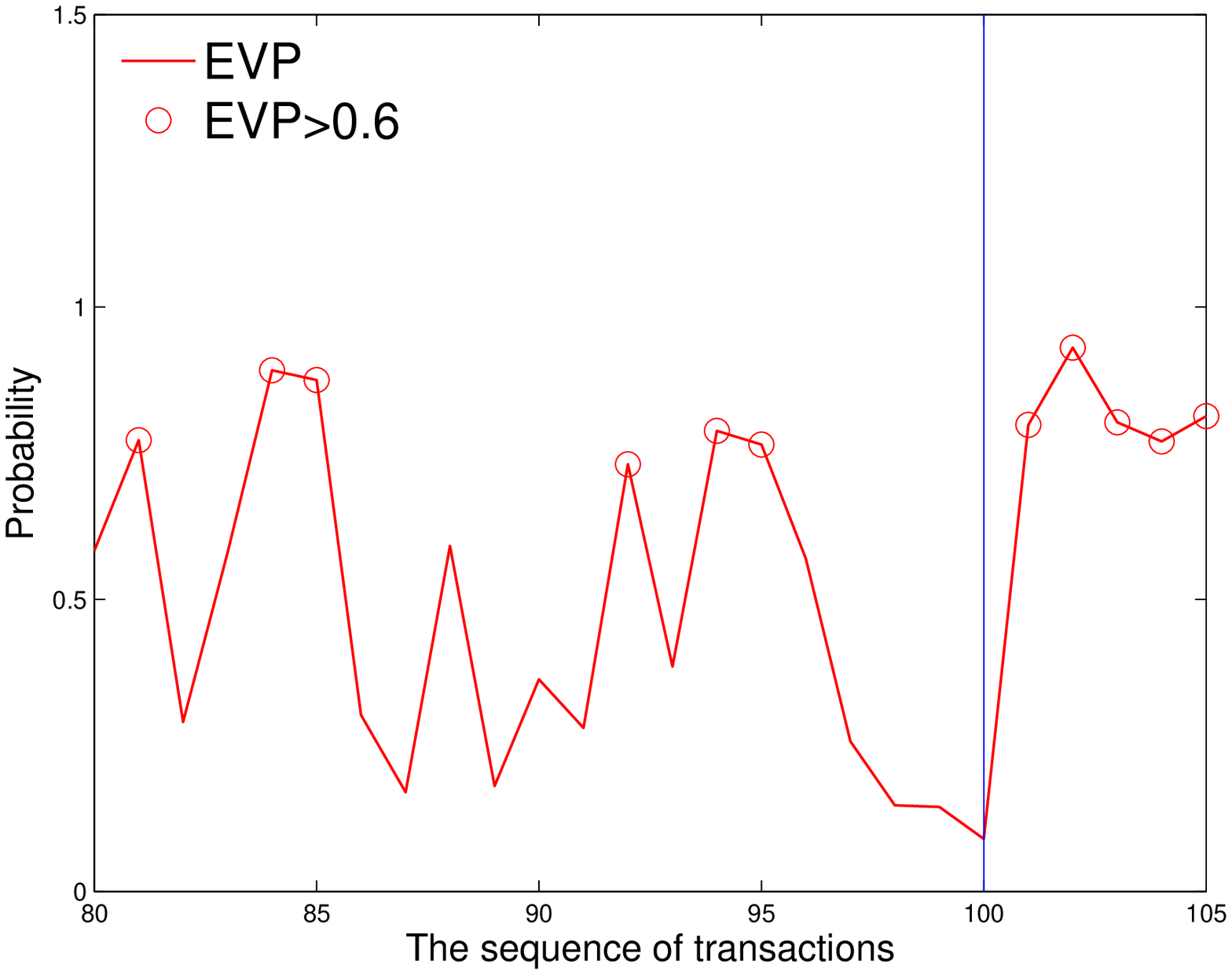} \cr
\includegraphics[scale=0.235]{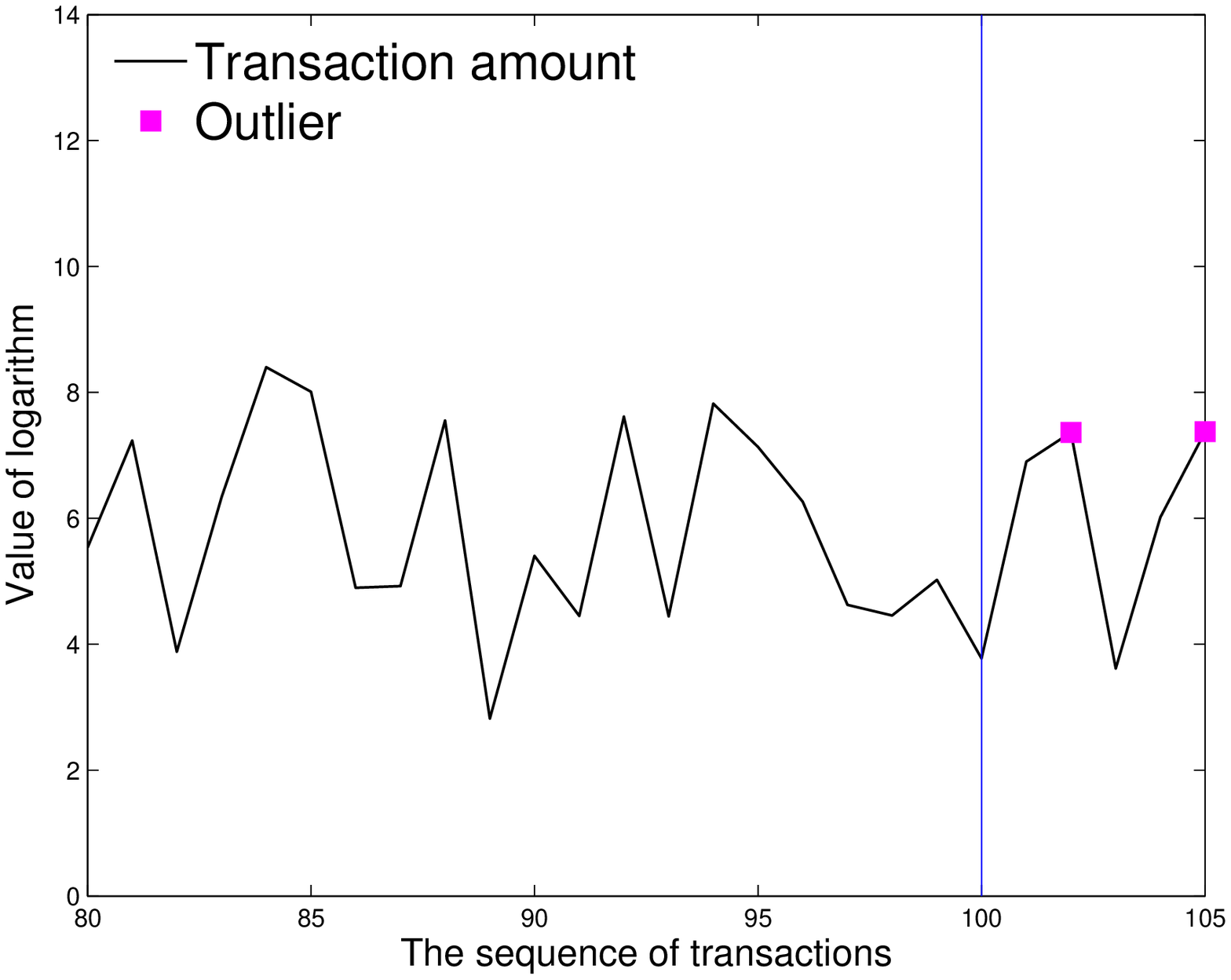} & 
\includegraphics[scale=0.235]{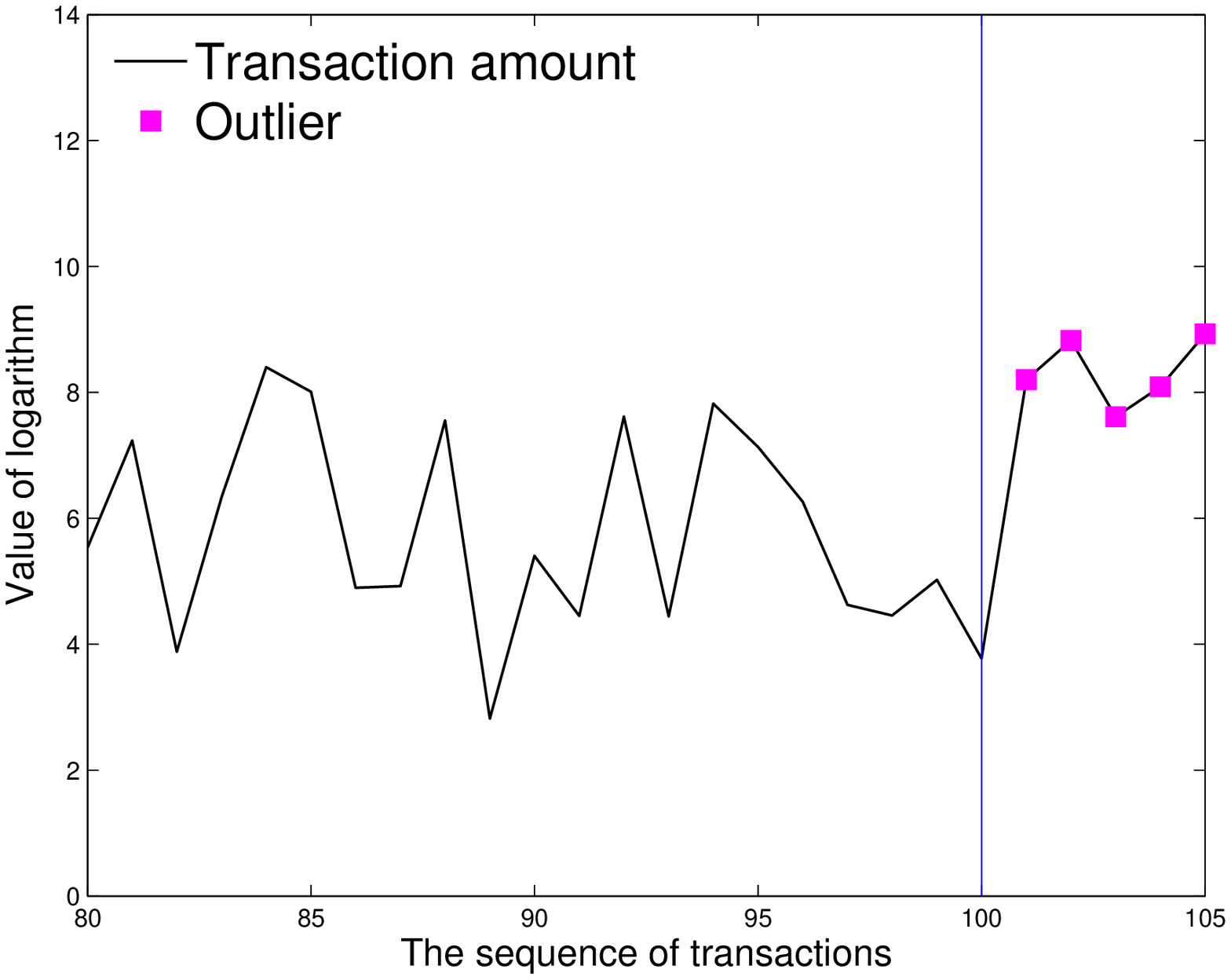} \cr
(a) Dataset $L$ &
(b) Dataset $F$
\end{tabular}
\caption{Outlier detection with GPs and EVP. Means and standard deviations of each transaction amounts are used to draw EVP and obtained by GPs.}
\label{fig:result6}
\end{figure}

Figs. \ref{fig:result5} and \ref{fig:result6} show the results of outlier detection for $L$ and $F$ using not the standard deviations but extreme-value probability (EVP) with the threshold $\theta_{EV} =0.6$. The red circles in the upper plots represent EVPs (i.e., $P_{EV}$) larger than 0.6, and the corresponding testing data are indicated as magenta squares in the lower plots. In Fig. \ref{fig:result5}, the distribution of testing data is supposed to be one-sided standard Gaussian having the estimated mean and standard deviation from the AR(5) model. The distribution of testing data in Fig. \ref{fig:result6} is from GPs. In these cases, we found that EVP gives a more reliable result because not only did both $AR(5)$ and GPs identify the fraudulent transactions in dataset $F$ as fraud more accurately than the +1 SD method, but also identified the legitimate transactions in dataset $L$ as normal more accurately than the +1 SD method. However, since there are some data such that $P_{EV}>0.6$ in a legitimate transaction sequence (i.e., on the left side of the vertical blue line in the upper plots), we can assume that it still gives rise to false-positive errors for the large amount transactions that are sometimes generated by a legitimate consumer.

Therefore, we consider the association rule and adjacency matrix, as well as $AR(5)$ and GPs, which handle the movement pattern of a consumer path, and determine the difference in the transaction pattern distribution of legitimate and fraudulent transactions. Then, we propose a optimal threshold for classifying them.

\subsection{Confidence on the transaction amount and region}

\begin{figure*}[t!]
\begin{center}
\begin{tabular}{m{1cm}|m{3.2cm}m{3.2cm}m{3.2cm}m{3.2cm}}
\hline
 &
\small{$x$-axis: Association rule} &
\small{$x$-axis: Association rule} &
\small{$x$-axis: Adjacency matrix} &
\small{$x$-axis: Adjacency matrix} \\

 &
\centering \small{$y$-axis: GPs} &
\centering \small{$y$-axis: $AR(5)$} &
\centering \small{$y$-axis: GPs} &
\centering \small{$y$-axis: $AR(5)$} \cr 

\hline
\centering \small{Dataset $L$}
 &
\includegraphics[scale=0.18]{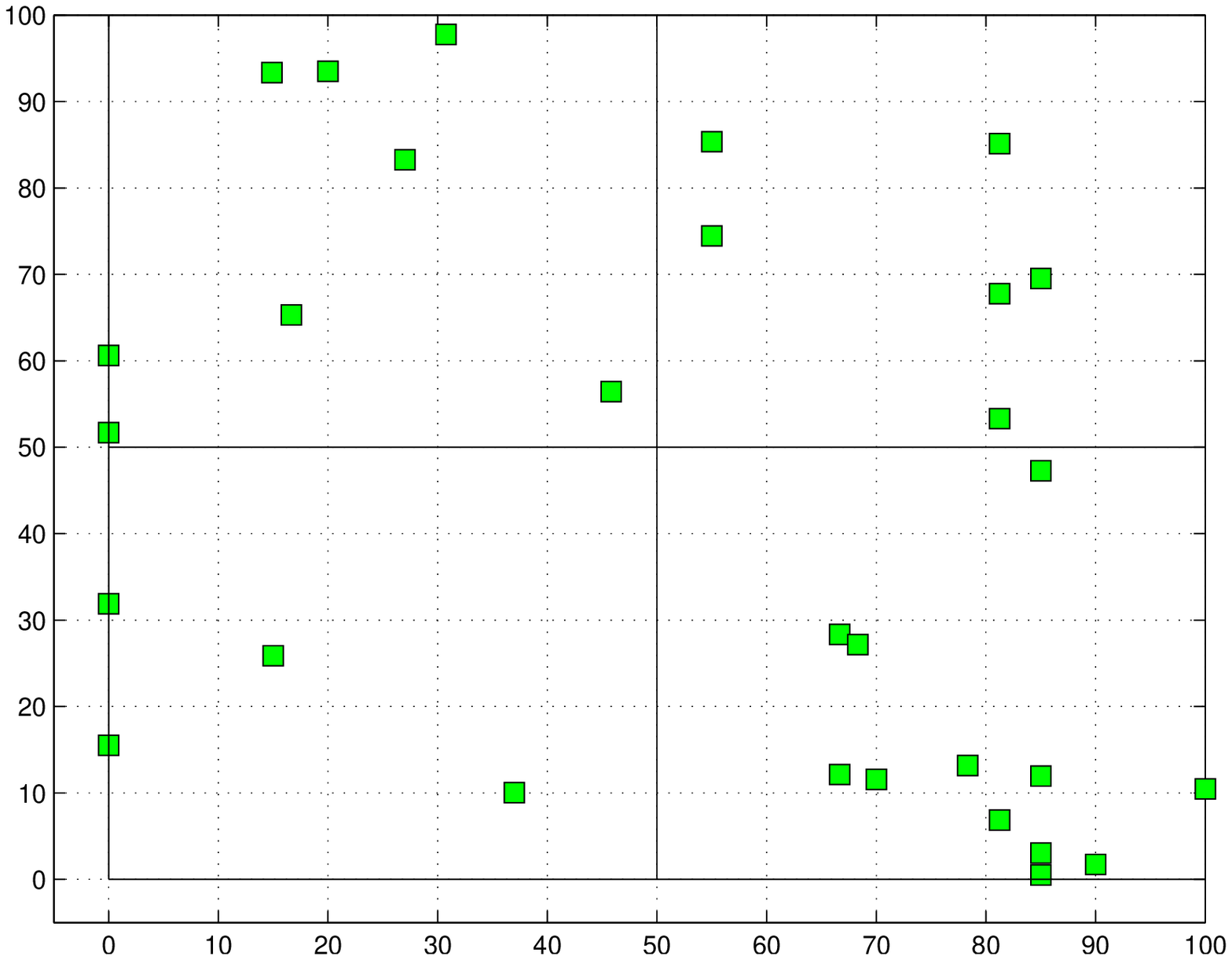} & 
\includegraphics[scale=0.18]{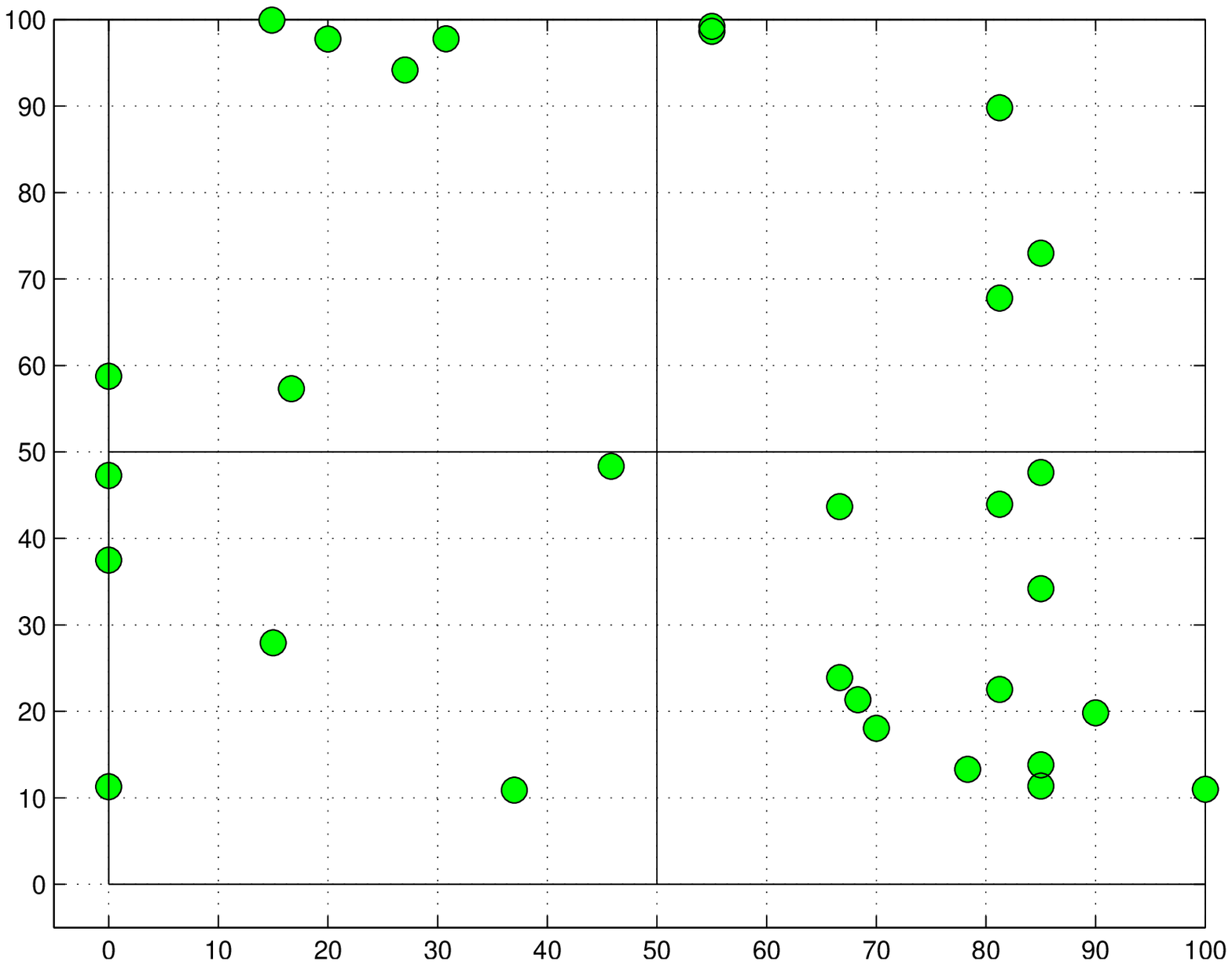} & 
\includegraphics[scale=0.18]{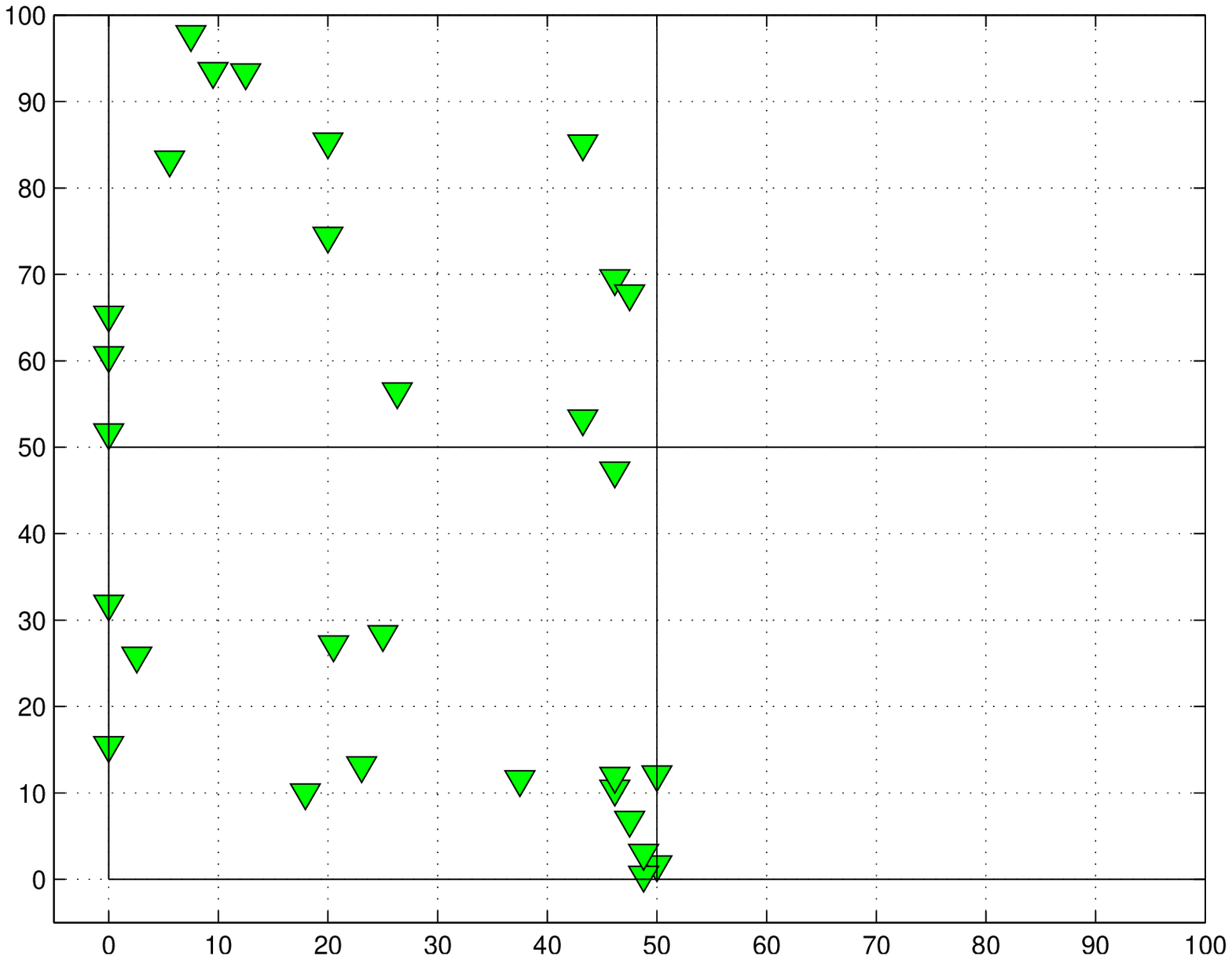} & 
\includegraphics[scale=0.18]{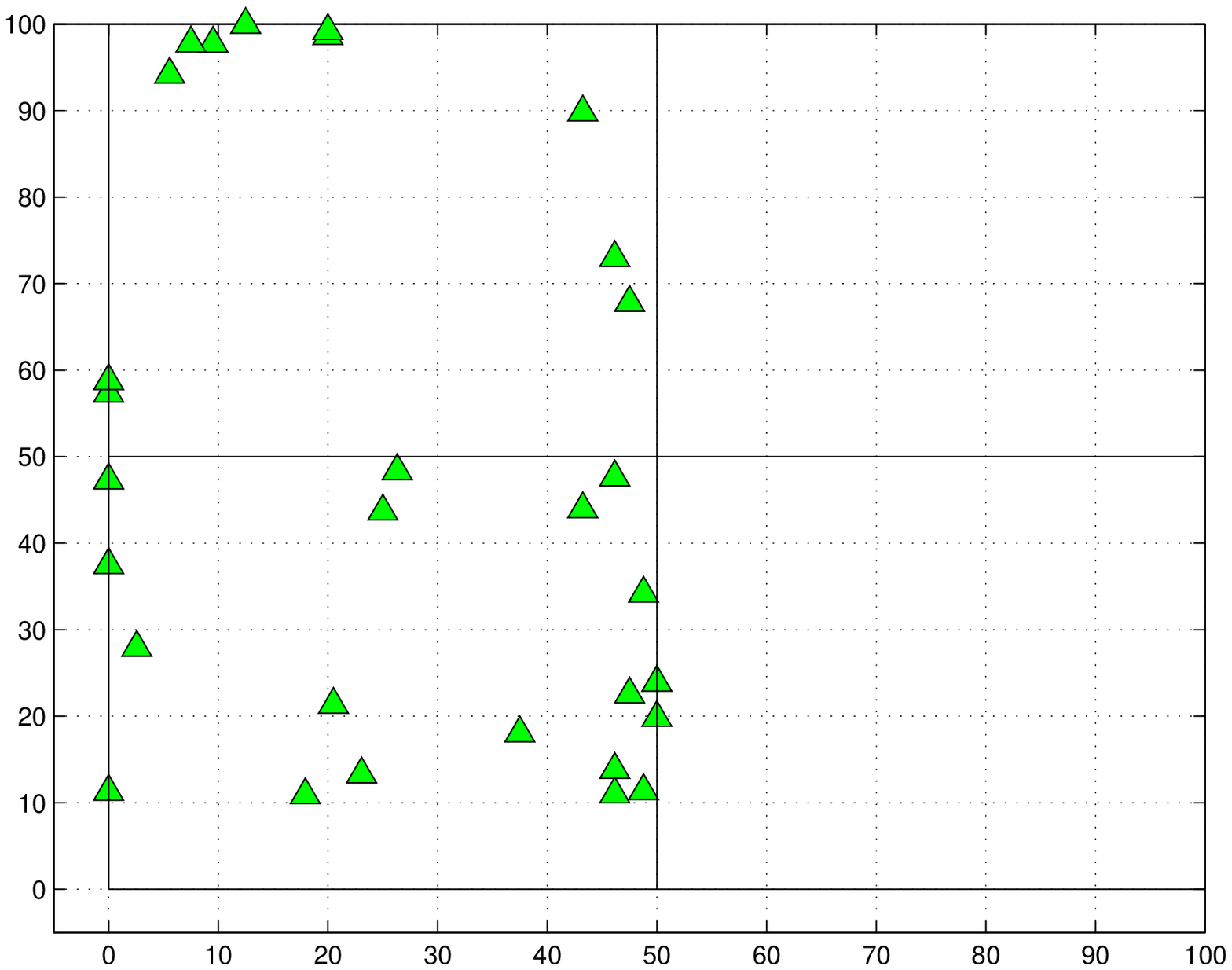} \cr

\centering \small{Dataset $F$}
 &
\includegraphics[scale=0.18]{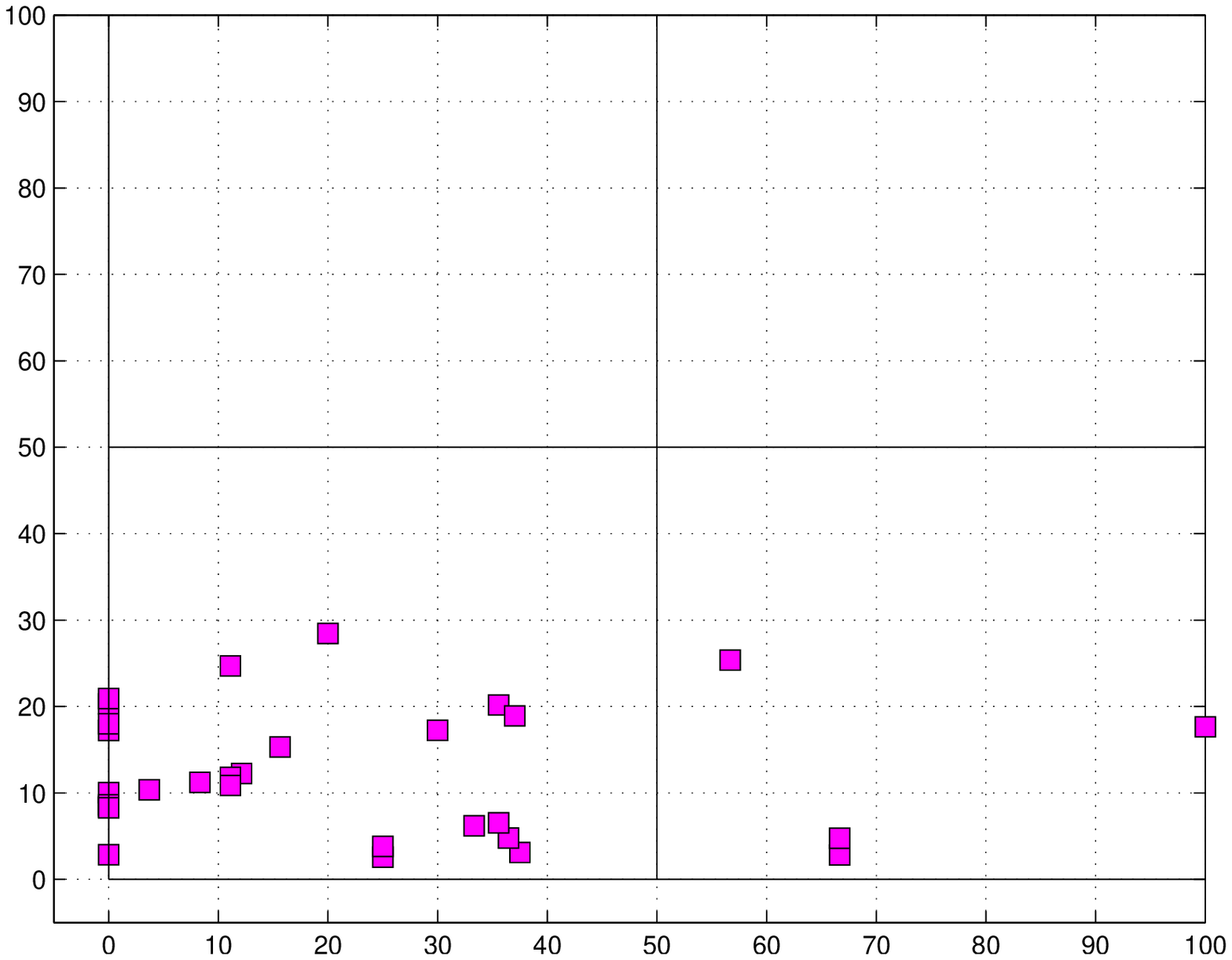} & 
\includegraphics[scale=0.18]{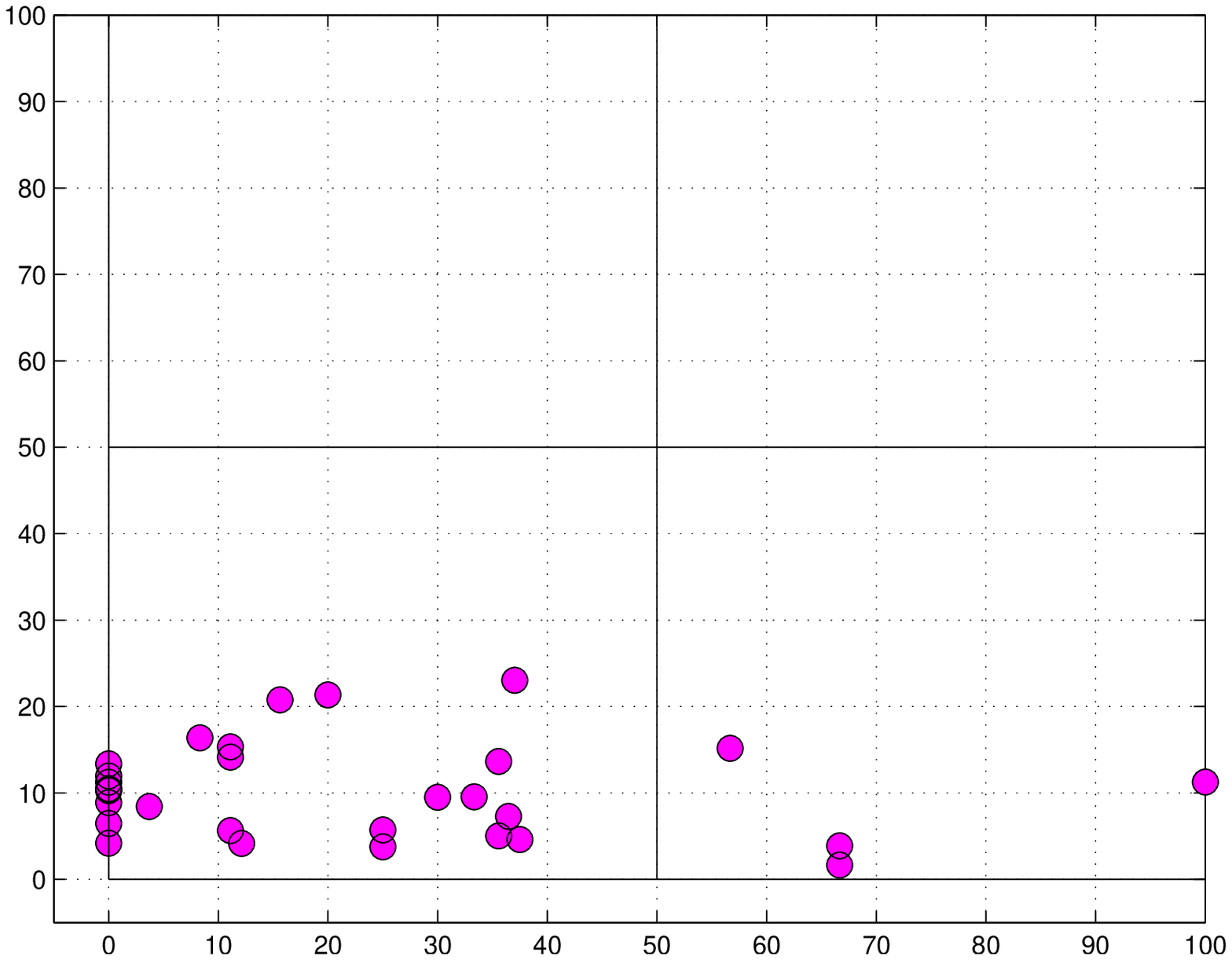} & 
\includegraphics[scale=0.18]{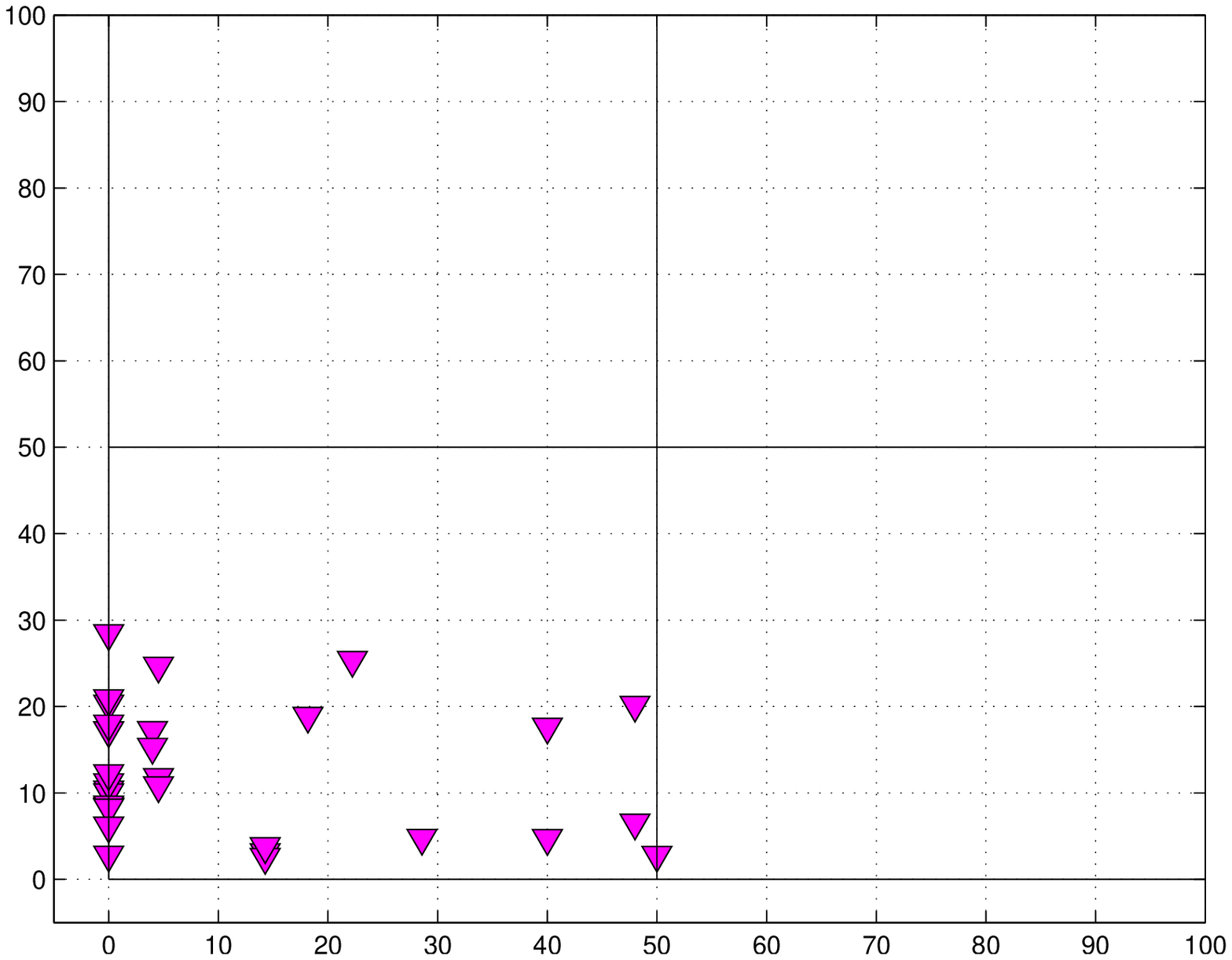} & 
\includegraphics[scale=0.18]{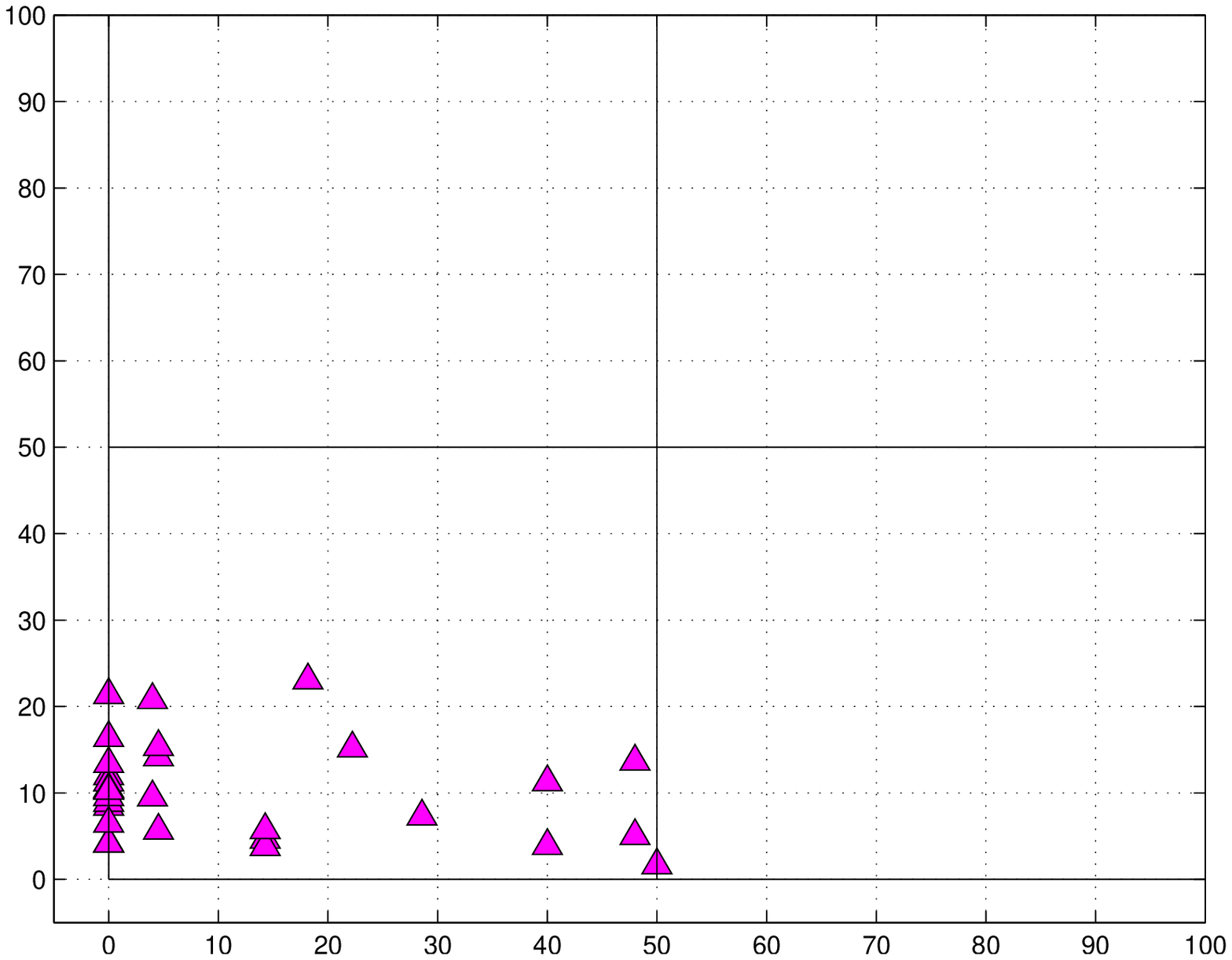} \cr
\hline
\end{tabular}
\caption{Comparison of four methods of classifying 30 transaction data. We first used the four methods to observe the distribution of transactions in datasets $L$ and $F$. Dataset $L$, which consists of legitimate transactions, is indicated as green markers and dataset $F$, which consists of fraudulent transactions, is indicated as magenta markers. We distinguished the methods by using different marker shapes for each. The markers in the plots of methods with the adjacency matrix are biased on the left side of the $xy$-plane.}
\label{fig:result7}
\end{center}
\end{figure*}

The confidence score for the transaction amount uses the cumulative distribution function (CDF) of Gaussian distribution, which has mean and standard deviations estimated from $AR(5)$ or GPs. In the case of Gaussian distribution with 0 mean and 1 variance (i.e., standard normal distribution), if the random variable of a newly occurred transaction amount is close to zero, (i.e., the average payment), we can understand that it follows the transaction pattern. On the other hand, if the random variable of a transaction amount is negative in standard normal distribution, we can ignore that event due to the small transaction amount, which is not threatening. In another case, when a random variable of the amount is positive in the distribution, we should pay attention to the event, since it represents that a large amount transaction exceeds the existing pattern has occurred. In order to express this possible threat as a numerical value, we define the confidence score for the transaction amount as ``1-CDF". That is, when the confidence score is higher, a smaller transaction amount will not cause concern, and when the confidence score is lower, a relatively large transaction amount that exceeds the pattern of the existing customer's transactions will be seen as threatening.

The confidence score for the transaction region is defined as the result value of the association rule analysis and the conditional probability of the adjacency matrix method.

Now, we can draw two attributions in a two-dimensional plane by placing the confidence score of the transaction region on the $x$ coordinate and that of the transaction amount on the $y$ coordinate. For each of 6 $L$ and $F$ datasets with a total of 30 transaction data (i.e., 30 testing data), we drew a simple distribution of the confidence score using the transaction amount and region data. Fig. \ref{fig:result7} shows 30 legitimate transaction data indicated by green points in the upper part and 30 fraudulent transaction data indicated by magenta points in the lower part. As shown in Fig. \ref{fig:result7}, the distribution of the experimental results for dataset $L$ is spread evenly, but in the case of dataset $F$ it is confirmed that the confidence scores are gathered at $x$ and $y$ coordinate values of less than 50. In particular, when the adjacency matrix is used for the confidence score for the transaction region (right side of Fig. \ref{fig:result7}), the $x$ coordinate value is biased at less than 50. This means that the association rule that considers the correlation between a variety of regions is a more appropriate method for processing the movement pattern of users than is the adjacency matrix, which deals with only the prior location.

\begin{figure}[h]
\begin{center}
\begin{tabular}{m{1cm}|m{3.2cm}m{3.2cm}}
\hline
 &
\centering Method 1 &
\centering Method 2 \cr
 &
\small{$x$-axis: Association rule} &
\small{$x$-axis: Association rule} \cr

 &
\centering \small{$y$-axis: GPs} &
\centering \small{$y$-axis: $AR(5)$} \cr 

\hline
\centering \small{Dataset $L$}
 &
\includegraphics[scale=0.18]{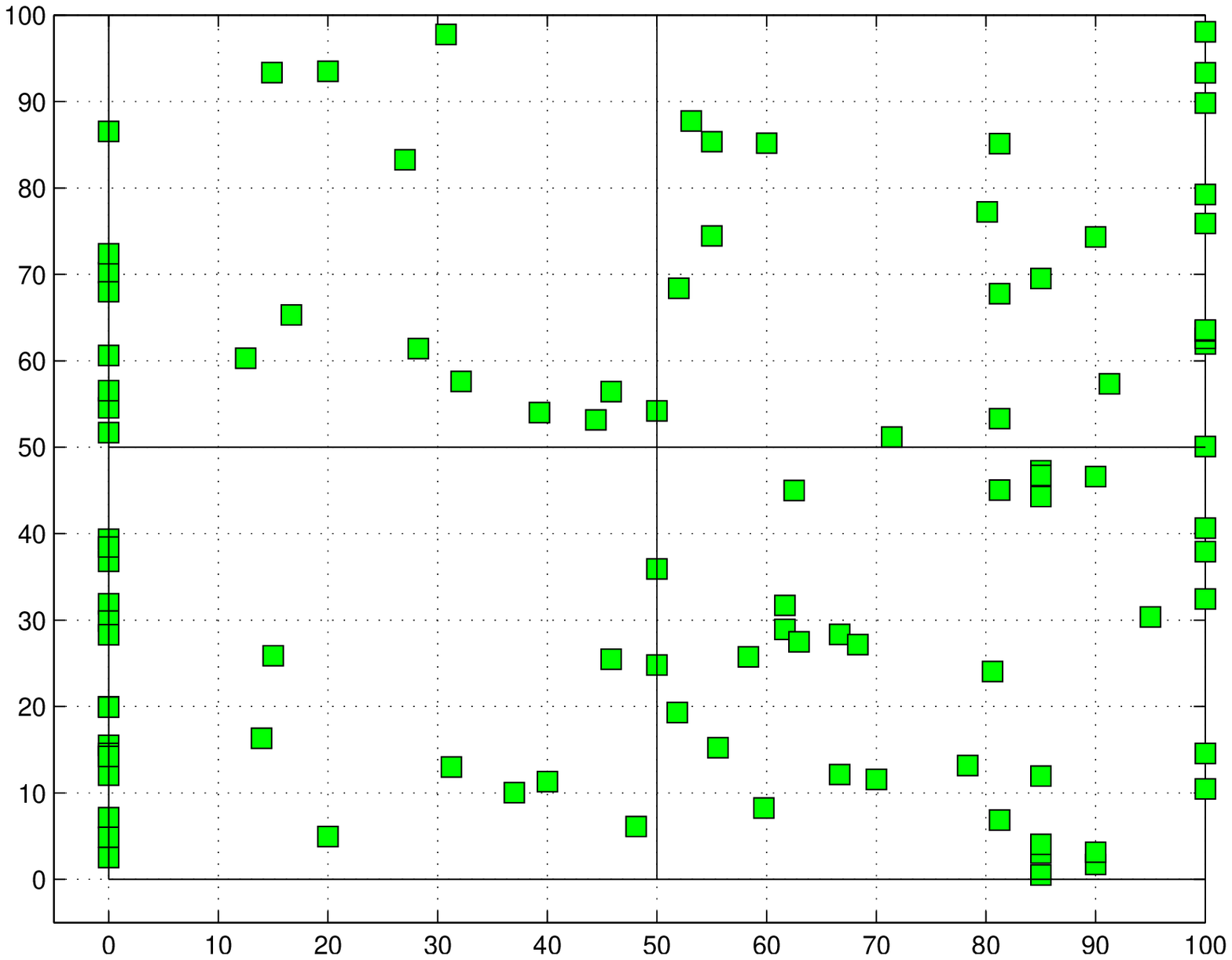} & 
\includegraphics[scale=0.18]{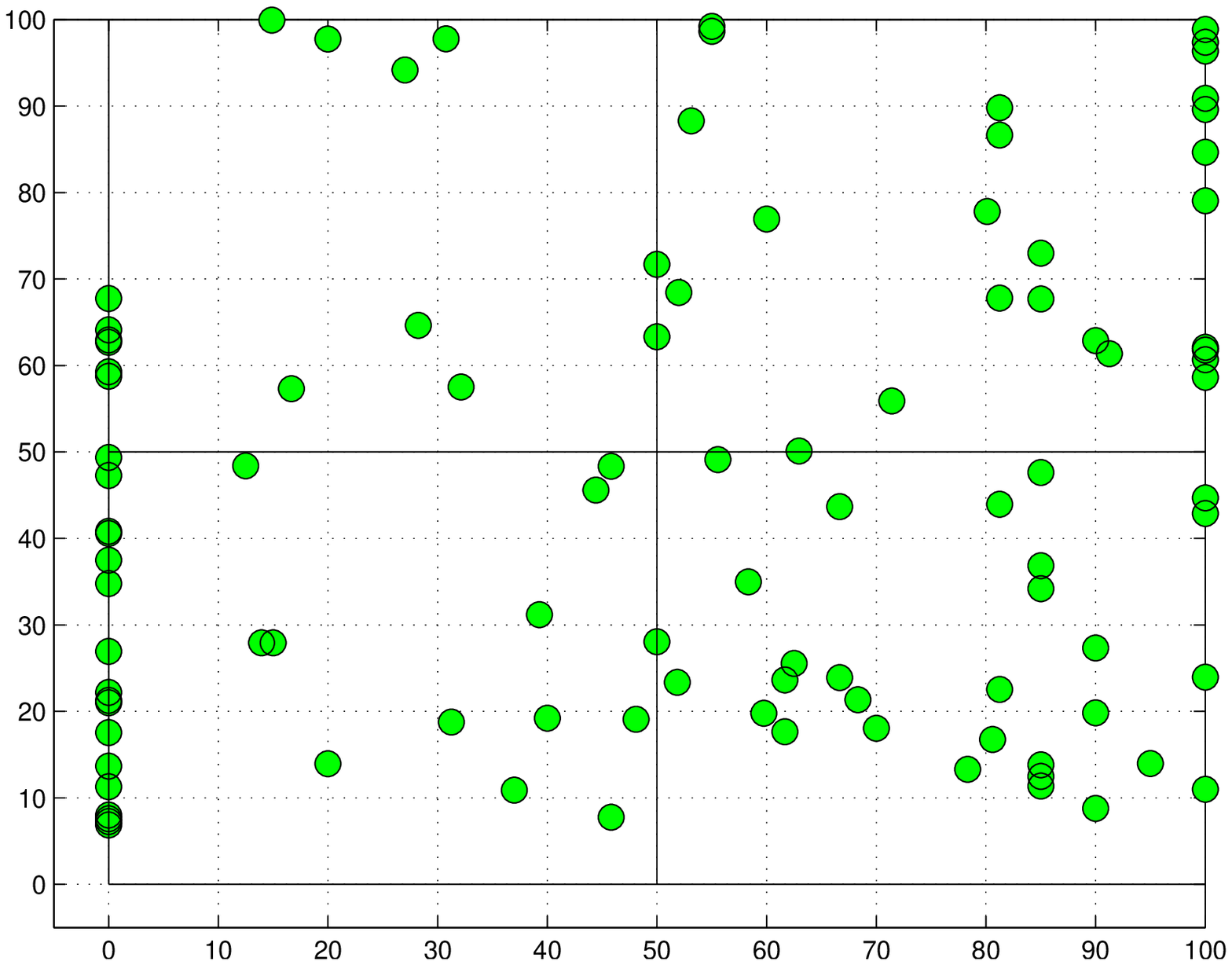}  \cr

\centering \small{Dataset $F$}
 &
\includegraphics[scale=0.18]{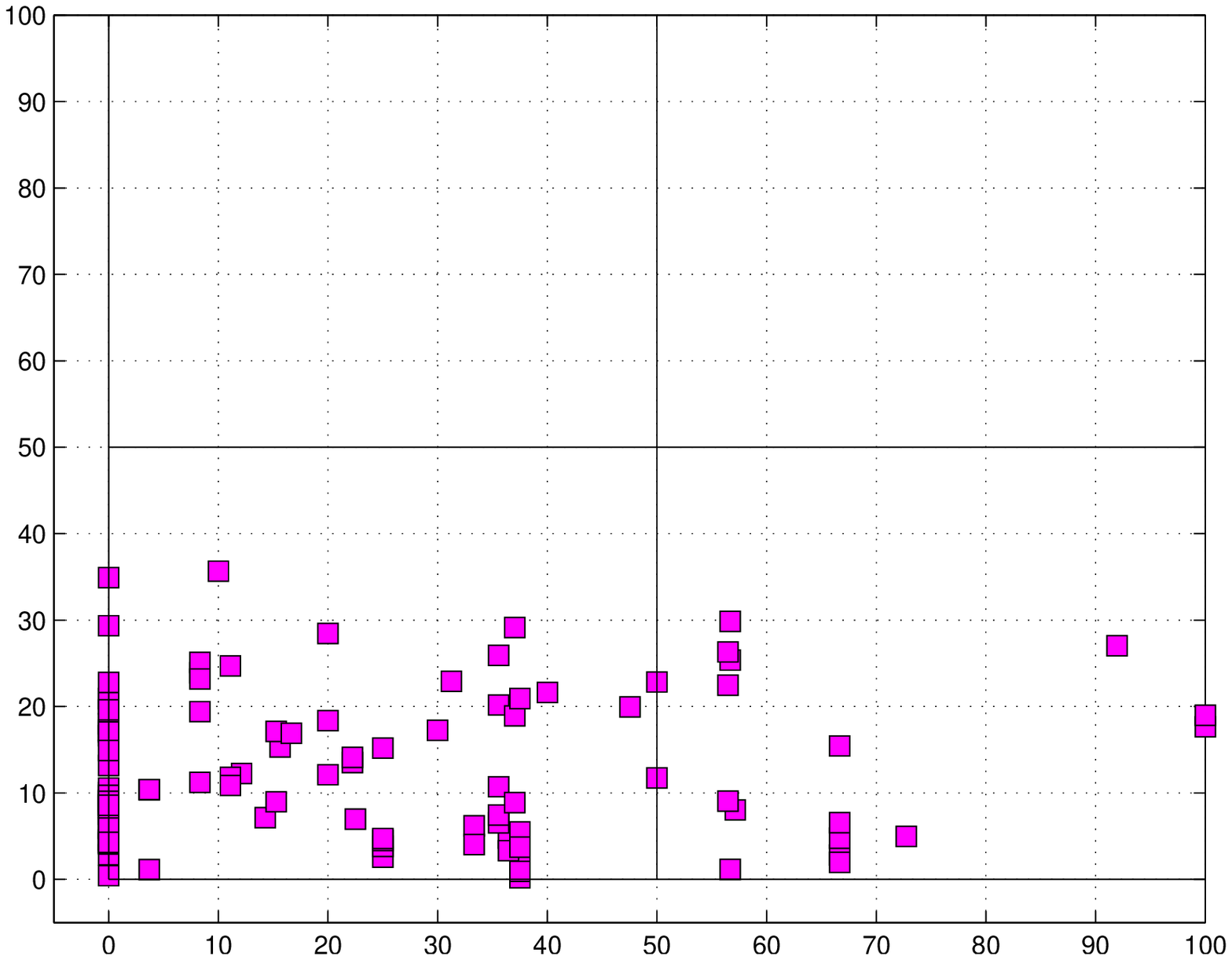} & 
\includegraphics[scale=0.18]{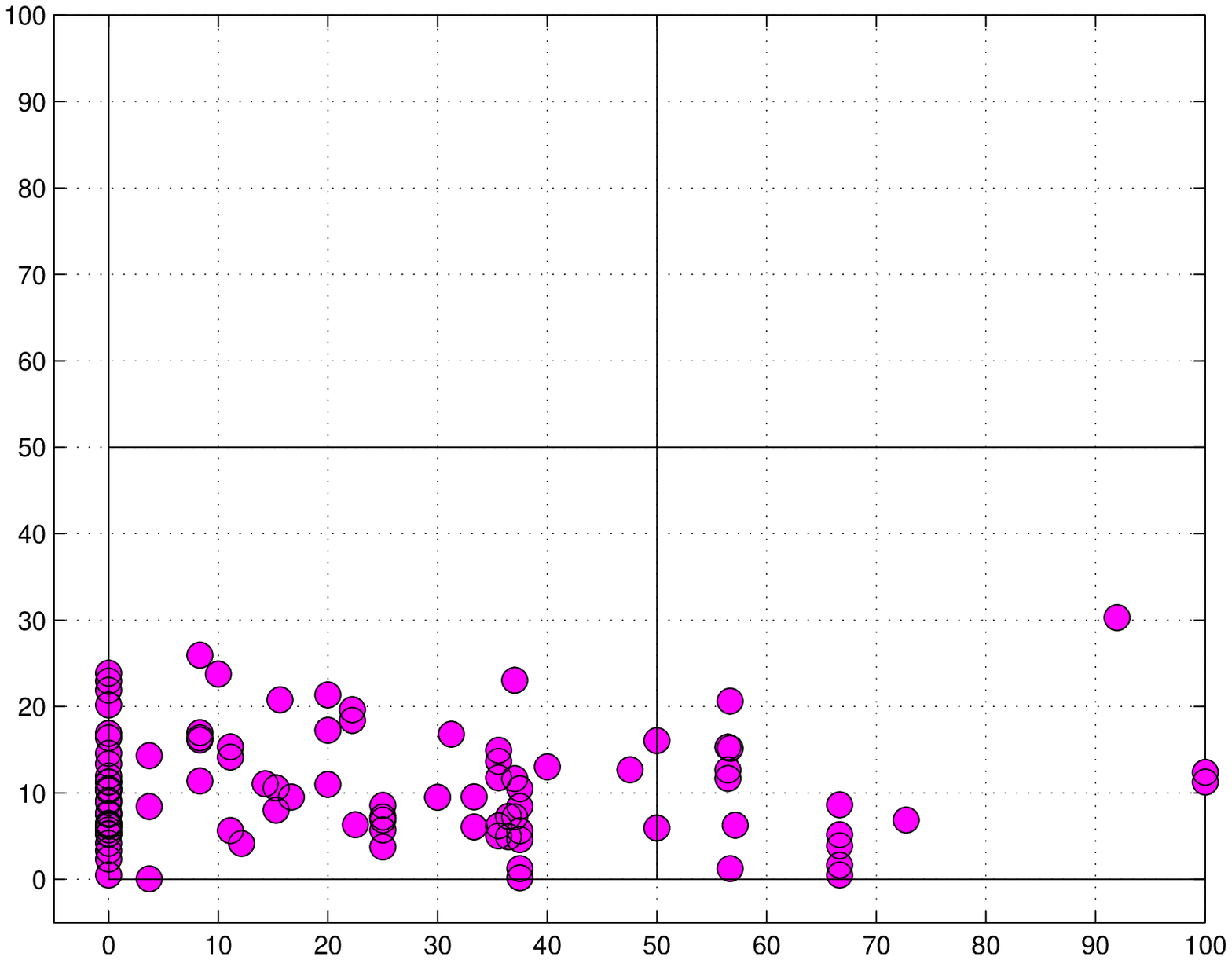} \cr
\hline
\end{tabular}
\caption{Comparison of two methods for classifying 100 transaction data. We chose the two better methods to observe the distribution of transactions in datasets $L$ and $F$ from Fig. \ref{fig:result7}. The legitimate transactions are evenly spread in various location while the fraudulent transactions are gathered around the origin of the coordinates.}
\label{fig:result8}
\end{center}
\end{figure}

Thus, we constrain the method for finding the confidence score of the transaction region to the association rule and not allow the adjacency matrix. With the association rule to obtain the confidence score of the transaction region, Method 1 uses GPs, and Method 2 uses $AR(5)$ to obtain that of the transaction amount. For each of 20 $L$ and $F$ datasets, a total 100 transaction data (i.e., 100 testing data), the comparison of the two methods is plotted in Fig. \ref{fig:result8}. As in Fig. \ref{fig:result7} the $F$ data are gathered at $x$ and $y$ coordinate values of less than 50. However, since the distribution of the confidence score for the $L$ data, indicated by circles in the upper part of the figure, has spread in various places, it is difficult to separate the dataset $F$ from the mixed dataset with $L$ and $F$ completely.

Therefore, even allowing some errors, we need to find a threshold $\theta_{x,y}$($=x<\theta$ and $y< \theta$) that maximizes the accuracy and minimizes the error rate, where the accuracy rate is defined as the ratio of the number of correctly predicted transaction to the total number of transaction data based on a given threshold. The errors are divided into false-positive and false-negative. The false-negative error rate is the ratio of the number of transactions that failed to identify a fraudulent transaction to the number of $F$ data, and the false-positive error rate is the ratio of the number of transactions that raised a false alarm for a legitimate transaction to the number of $L$ data.

For example, based on a threshold $\theta_{x,y}=30$, the accuracy at $\theta_{x,y}=30$ is the sum of the number of markers in dataset $L$ where $x\geq30$ or $y\geq30$ and the number of markers in dataset $F$ where $x<30$ and $y<30$. The false-positive error is the number of markers in dataset $L$ where $x<30$ and $y<30$, and the false-negative error is the number of markers in dataset $F$ where $x\geq30$ or $y\geq30$.

\begin{table*}[t!]
\begin{center}
    \begin{tabular}{|cc|ccccc|}
    \hline
    ~        & ~              & $\theta_{x,y}=10$ & $\theta_{x,y}=20$ & $\theta_{x,y}=30$ & $\theta_{x,y}=40$ & $\theta_{x,y}=50$ \\ \hline
    Method 1 & Accuracy       & 115(.58)          & 129(.65)          & 144(.72)          & \textbf{160(.80)} & 159(.80)   \\
    Method 2 & ~              & 115(.58)           & 135(.68)          & 146(.73)          & \textbf{161(.81)}& 153(.77)   \\ \hline
    Method 1 & False-positive & 3        & 10        & 14        & \textbf{20}     & 23      \\
    ~        & False-negative & 82       & 61        & 42        & \textbf{20}     & 18      \\
    Method 2 & False-positive & 4        & 7         & 14        & \textbf{19}     & 29      \\
    ~        & False-negative & 81       & 58        & 40        & \textbf{20}     & 18      \\ \hline
    \end{tabular}
    \caption {The accuracy and error rate of each method. Method 1 is a combination method of association rule and GPs (i.e., rectangular markers in Figs. \ref{fig:result7} and \ref{fig:result8}). Method 2 is a combination method of association rule and $AR(5)$ (i.e., circle markers in Figs. \ref{fig:result7} and \ref{fig:result8}). We determined the optimal thresholds with the highest accuracy and the lowest error rates, indicated in bold type.}
    \label{tab:result}
\end{center}
\end{table*}

As a result, $\theta_{x,y}=40$ for Method 1 and $\theta_{x,y}=50$ for Method 2 is the optimal threshold that maximizes the accuracy and minimizes the error rate, as shown in detail in Table \ref{tab:result}.

\section{Discussion}
\label{section:Sec6}
Until now, we gave the confidence scores to a customer's transaction amount and region according to the previous transaction patterns, and examined the characteristics of the distribution of the values by plotting them in a coordinate plane. As a result, it was not easy to discriminate completely the data as legitimate or fraudulent, but by setting the appropriate threshold we can obtain the optimal solution. In spite of these contributions, a discussion about the false-positive errors and false-negative errors that occurred in this experiment is required. The reasons for the errors and countermeasures that can be considered are as follows:

\begin{enumerate}
\item The fraudulent transaction amount induced from Table 4 of \citet{ref3} was not very significantly higher than the legitimate transaction amount. Some samples of fraudulent transaction data contain a normal transaction amount that is not larger than the legitimate transaction amount.

However, this may not be a problem in real-life transactions because perpetrators attempt to use stolen cards or duplicate cards quickly to maximize the amount of fraudulent transactions so that they can complete before the fraud comes to light.

\item Immediately after a legitimate transaction involving the payment of a large amount of money occurs, the estimated amount of the next transaction is increased in accordance the amount of the previous transaction. Therefore, a fraudulent transaction involving a relatively small amount is not identified, and hence, a false-negative error occurs.

This issue can be resolved by reducing the length of the testing data, the number of transactions processed to identify the fraud. We used 20 $B$ datasets as testing data in dataset $F$. Each the fraudulent set had five transactions (i.e., 5-length). In our experiment using dataset $F$, confidence scores of the first and second data of the fraudulent $B$ sets were generally lower than those of the third, fourth, and fifth data.

Fig. \ref{fig:result10} shows the confidence scores of 20 arbitrary fraudulent transaction data of the $B$ datasets obtained by GPs. They are plotted in ascending order. The confidence scores of the first $B$ data are represented by a yellow line, which is lower than 10. Confidence scores of the second data in the $B$ dataset are represented by an orange line, which is almost lower than 10. This means that if we constrained the number of testing data as 2, there would be less false-negative errors for the fraudulent transactions in GPs using a threshold $\theta_{x,y}=10$.

\begin{figure}[h!]
\begin{center}
\includegraphics[scale=0.6]{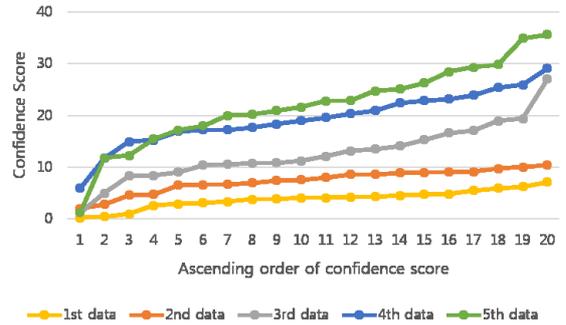}
\caption{Confidence scores of fraudulent transaction set $B$ in the order of testing data obtained by GPs. An arbitrary fraudulent transaction set $B$ has 5 transactions. For each of 20 fraudulent $B$ testing datasets, the confidence scores of the first data in the $B$ datasets are represented by orange dots and lines in ascending order. The second to fifth data are also represented by different colors. This figure shows that the first and second transactions of the $B$ datasets almost never exceeded a confidence score of 10. This demonstrates that the shorter the length of the testing data used, the lower is the false-negative error rate in not only GPs but also $AR(5)$.}
\label{fig:result10}
\end{center}
\end{figure}

\item Conversely, after a legitimate transaction involving the payment of a small amount of money occurs, the estimated amount of the next transaction is decreased in accordance the amount of the previous transaction. Therefore, a legitimate transaction involving a relatively large amount raises a false-alarm, and hence, gives rise to a false-positive error.

This issue can also be resolved by adding a certain transaction amount that may have occurred in a fraudulent transaction to the threshold of confidence scores.

\item Because the confidence score is 0 for the first visit location, many data are gathered on the $y$-axis, whether the transaction is legitimate or fraudulent.

For consumers whose lifestyle is fixed and regular, it seems that the cases where the data are clustered on the $y$-axis are usually fewer. In addition, in this experiment the region of a fraudulent transaction is obtained randomly from the region of a legitimate transaction. However, since in real life the region of a fraudulent transaction is completely independent of the region of a legitimate transaction, there may be more fraudulent transaction data near the $y$-axis.
\end{enumerate}

\section{Conclusion}
\label{section:Sec7}
This paper proposed methods that use the transaction patterns of a consumer to detect fraudulent card use.

For many reasons, such as changes in income, lifestyle, and place of abode, the consumption patterns of credit card users also change frequently. Therefore, it is difficult to detect a fraudulent card transaction when only certain fixed conditions have been set. In this paper, we present a method for learning automatically the consumption pattern of a customer, and detecting the fraudulent use of a card that has been duplicated or stolen based on the pattern.

This study also addressed 2-dimensional methods by not only dealing with the amount of transactions but also highlighting the patterns in the region in which a transaction occurs.


\bibliographystyle{model5-names}
\bibliography{eswa}

\end{document}